\documentclass[conference]{IEEEtran}

\IEEEoverridecommandlockouts

\usepackage{cite}
\usepackage{graphicx}
\usepackage{textcomp}
\usepackage{xcolor}
\def\BibTeX{{\rm B\kern-.05em{\sc i\kern-.025em b}\kern-.08em
    T\kern-.1667em\lower.7ex\hbox{E}\kern-.125emX}}

\usepackage{microtype}
\usepackage[export]{adjustbox}
\usepackage{subcaption}
\usepackage{booktabs} 
\usepackage{array}
\usepackage{enumitem}

\usepackage{tikz}
\usepackage{pifont}
\usepackage[utf8]{inputenc}
\usepackage{tabularx}
\usepackage{bm}
\usepackage{multirow}

\usepackage[nolist]{acronym}

\usepackage[intoc, english]{nomencl}
\makenomenclature

\usepackage{xspace}

\usepackage[font=small]{caption}

\usepackage[colorlinks=false,urlcolor=black,hidelinks]{hyperref}

\usepackage{amsmath, amsthm, amssymb}

\usepackage{algorithm}
\usepackage{algorithmicx}
\usepackage{algpseudocode}
\usepackage{graphics}
\newcommand{\model}{\Phi}

\newcommand{\inputaudio}{\mathbf{x}}

\newcommand{\advpairsset}{T}

\newcommand{\originalstate}{Y}
\newcommand{\adversarialstate}{Z}

\newcommand{\numadvpair}{N}
\newcommand{\audioframe}{x}

\newcommand{\word}{W}
\newcommand{\hmmstate}{w}

\newcommand{\nummodels}{M}
\newcommand{\modelindex}{m}
\newcommand{\poisonindex}{p}
\newcommand{\poisonframe}{\audioframe_\gamma}
\newcommand{\originalframe}{\audioframe_b}
\newcommand{\numpois}{P}

\newcommand{\indexone}{i}
\newcommand{\indextwo}{j}

\newcommand{\wordstateindex}{\kappa}

\newcommand{\poisonbudget}{r_p}
\newcommand{\maxiter}{K}

\newcommand{\numframes}{K}
\newcommand{\frameindex}{k}
\newcommand{\noise}{\sigma}
\newcommand{\numtimeframes}{T}
\newcommand{\timeframesindex}{t}

\newcommand{\substitutions}{S}
\newcommand{\insertions}{I}
\newcommand{\deletions}{D}
\newcommand{\numwords}{N}

\newcommand{\origspect}{O}
\newcommand{\poisonspect}{\Upsilon}
\newcommand{\diffspect}{D}
\newcommand{\audibleDiff}{\zeta}
\newcommand{\audibleDiffMinZero}{\zeta^*}
\newcommand{\audibleDiffNormalized}{\hat{\zeta}}
\newcommand{\hearingThresholds}{H}
\newcommand{\hearingThresholdsNormalized}{\hat{H}}
\newcommand{\hearingOffset}{\Lambda}

\newcommand\surrogateHMM{\mathcal{H}}
\newcommand\surrogateDNN{\mathcal{M}}
\newcommand\cleanDataset{\mathcal{C}}
\newcommand\poisonDataset{\mathcal{P}}

\renewcommand{\nomgroup}[1]{%
\ifthenelse{\equal{#1}{F}}{}{
\ifthenelse{\equal{#1}{G}}{\item[\textbf{Greek symbols}]}{%
\ifthenelse{\equal{#1}{R}}{\item[\textbf{Roman symbols}]}{}}}\bigskip
}

\nomenclature[F]{$\originalstate$}{Original state}
\nomenclature[F]{$\adversarialstate$}{Adversarial state}
\nomenclature[F]{$\numwordshmmstates$}{Number of states in a word}
\nomenclature[F]{$\word = \{\hmmstate_0, \hmmstate_1, \dots, \hmmstate_\numwordshmmstates\}$}{Word definition as HMM states}
\nomenclature[F]{$\audioframe$}{Audio frame}
\nomenclature[F]{$\audioframe_{(\originalstate, \adversarialstate)}$}{Audio frame with benign state~$\originalstate$ and adversarial state~$\adversarialstate$}
\nomenclature[F]{$\advpairsset = [\audioframe_{(\originalstate_0, \adversarialstate_0)}, \audioframe_{(\originalstate_1, \adversarialstate_1)}, \dots, \audioframe_{(\originalstate_\numadvpair, \adversarialstate_\numadvpair)}]$}{Sequence of adversarial frames}
\nomenclature[F]{$\nummodels$}{Number of substitution models}
\nomenclature[F]{$\model$}{Model}
\nomenclature[F]{$\poisonframe$}{Poison frame}
\nomenclature[F]{$\originalframe$}{Original frame}
\nomenclature[F]{$\targetframe$}{Target frame}
\nomenclature[F]{$\numpois$}{Number of poisons}

\newcommand{\eg}{e.\,g.,\xspace}
\newcommand{\ie}{i.\,e.,\xspace}

\newcommand{\etal}{et~al.\@\xspace}

\newcommand\norm[1]{\left\lVert#1\right\rVert}

\def\sys{\textsc{Veno\-Mave}\xspace}

\def\Psycho{Psychoacoustic}
\def\psycho{psychoacoustic}

\def\netTwoLayer{$DNN_2$}
\def\netTwoLayerPlus{$DNN_{2+}$}
\def\netThreeLayer{$DNN_3$}
\def\netThreeLayerPlus{$DNN_{3+}$}

\def\netTwoLayerBold{$\bm{DNN_2}$}
\def\netTwoLayerPlusBold{$\bm{DNN_{\bm{2+}}}$}
\def\netThreeLayerBold{$\bm{DNN_3}$}
\def\netThreeLayerPlusBold{$\bm{DNN_{3+}}$}

\newcolumntype{C}[1]{>{\hsize=#1\hsize\centering\arraybackslash}X}
\newcolumntype{L}[1]{>{\hsize=#1\hsize\arraybackslash}X}
\newcolumntype{R}[1]{>{\hsize=#1\hsize\RaggedLeft\arraybackslash}X}

\newcommand{\cmark}{\ding{51}}%
\newcommand{\xmark}{\ding{55}}%
\begin{acronym}
    \acro{ASR}{\emph{Automatic Speech Recognition}}
    \acro{CTC}{\emph{Connectionist Temporal Classification}}
    \acro{DCT}{\emph{Discrete Cosine Transformation}}
    \acro{DFT}{\emph{Discrete Fourier Transform}}
    \acro{DNN}{\emph{Deep Neural Network}}
    \acro{DNN-HMM}{\emph{Deep Neural  Network - Hidden  Markov  Model}}
    \acro{GMM}{\emph{Gaussian Mixture Model}}
    \acroplural{GMM}[GMMs]{\emph{Gaussian Mixture Models}}
    \acro{HMM}{\emph{Hidden Markov Model}}
    \acro{MFCC}{\emph{Mel Frequency Cepstral Coefficient}}
    \acroplural{MFCC}[MFCCs]{\emph{Mel Frequency Cepstral Coefficients}}
    \acro{RIR}{\emph{Room Impulse Response}}
    \acro{SNRseg}{\emph{Segmental Signal-to-Noise Ratio}}
    \acro{SNR}{\emph{Signal-to-Noise Ratio}}
    \acro{WER}{\emph{Word Error Rate}}
    \acro{WSJ}{\emph{Wall Street Journal}}
\end{acronym}

\begin{document}

\title{\sys{}: Targeted Poisoning\\Against Speech Recognition}

\author{\IEEEauthorblockN{Hojjat Aghakhani\IEEEauthorrefmark{1},
Lea Schönherr\IEEEauthorrefmark{2},
Thorsten Eisenhofer\IEEEauthorrefmark{3},
Dorothea Kolossa\IEEEauthorrefmark{4},
Thorsten Holz\IEEEauthorrefmark{2},\\
Christopher Kruegel\IEEEauthorrefmark{1}, and Giovanni Vigna\IEEEauthorrefmark{1}
}
\IEEEauthorblockA{\IEEEauthorrefmark{1}University of California, Santa Barbara \IEEEauthorrefmark{2}CISPA Helmholtz Center for Information Security \\\IEEEauthorrefmark{3}Ruhr University Bochum \IEEEauthorrefmark{4}Technische Universität Berlin}

\IEEEauthorblockA{\IEEEauthorrefmark{1}\{hojjat, chris, vigna\}@cs.ucsb.edu \IEEEauthorrefmark{2}\{schoenherr, holz\}@cispa.de \IEEEauthorrefmark{3}thorsten.eisenhofer@rub.de \IEEEauthorrefmark{4}dorothea.kolossa@tu-berlin.de}
}

\maketitle

\begin{abstract}


Despite remarkable improvements, automatic speech recognition is susceptible to adversarial perturbations. Compared to standard machine learning architectures, these attacks are significantly more challenging, especially since the inputs to a speech recognition system are time series that contain both acoustic and linguistic properties of speech. Extracting all recognition-relevant information requires more complex pipelines and an ensemble of specialized components. Consequently, an attacker needs to consider the entire~pipeline.

In this paper, we present \sys{}, the first training-time poisoning attack against speech recognition. Similar to the predominantly studied evasion attacks, we pursue the same goal: leading the system to an incorrect and attacker-chosen transcription of a target audio waveform. In contrast to evasion attacks, however, we assume that the attacker can only manipulate a small part of the \emph{training data} without altering the target audio waveform at runtime.
We evaluate our attack on two datasets: TIDIGITS and \emph{Speech Commands}. When poisoning less than 0.17\,\% of the dataset, \sys{} achieves attack success rates of more than 80.0\,\%, \emph{without} access to the victim's network architecture or hyperparameters. In a more realistic scenario, when the target audio waveform is played over the air in different rooms, \sys{} maintains a success rate of up to~73.3\,\%. Finally, \sys{} achieves an attack transferability rate of 36.4\,\% between two different \emph{model architectures}.

\end{abstract}



\section{Introduction}
\label{sec:intro}
\begin{figure*}[t]
    \centering
    \includegraphics[trim=10 5 0 5, clip, width=\textwidth]{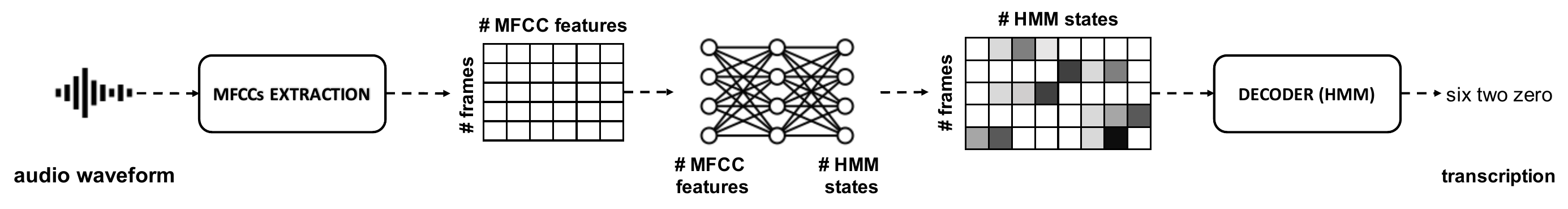}
    \vspace{-1.4em}
    \caption{\textbf{Overview of a state-of-the-art hybrid ASR system}. The ASR system is composed of two main components: The neural network acts as an acoustic model, and the decoder employs a \ac{HMM} to generate the transcription. The \ac{HMM} mainly describes the language grammar, a phonetic-based word description of all words, and context-dependencies of phonetic units and words.}
    \label{fig:asr_overview}
    \vspace{-0.8em}
\end{figure*}

Digital voice assistants are ubiquitous, whether at our homes, in our cars, or on our smartphones. 
Forecasts predict that by 2024, the number of digital voice assistants will surpass the world's population with more than 8 billion devices~\cite{vailshery-20-numberassistants}.
While there is a constant effort in improving their built-in \ac{ASR}, prior research~\cite{carlini2018audio, schonherr2018adversarial, abdullah-20-sok} has demonstrated that \ac{ASR} systems are susceptible to adversarial examples, i.e., malicious audio inputs that trigger a misclassification at \emph{runtime}. 
Such evasion attacks are a well-studied phenomenon and have been demonstrated to work for various domains~\cite{goodfellow2014explaining, ilyas-19-adversarial}, including speech recognition~\cite{carlini2016hidden, carlini2018audio, schonherr2018adversarial}.
In contrast, attacks \textit{during training} of \ac{ASR}, so-called \emph{poisoning attacks}~\cite{goldblum-21-poisonsurvey, biggio2012poisoning, zhu2019transferable}, have not been studied yet~\cite{abdullah-20-sok}.
Unlike evasion attacks, poisoning attacks  compromise the training data and cause misclassification of \emph{unaltered} inputs during inference. 
Consequently, such an attack is hard to detect, as the training data is usually not released with the model.

Poisoning attacks are enabled by the massive amounts of data needed to train machine learning models: State-of-the-art \ac{ASR} systems require thousands or even millions of samples, which makes it infeasible to manually verify the training set. It is common practice to collect datasets from potentially untrustworthy sources (\eg through crowd-sourcing or using open-source repositories).
Even more problematic are privacy-preserving training approaches like federated learning, which make it even easier to compromise the training process~\cite{bagdasaryan-20-fedpois, bhagoji-19-poisfed}. By design, the training data does not leave the client and can therefore not be verified. This property can be leveraged by a malicious party to feed the model with poisoned~data. 
Acknowledging these concerns, a recent survey of 28 industry organizations found that industry practitioners ranked \textit{data poisoning} as the most serious threat to ML systems~\cite{kumar2020adversarial}, emphasizing that poisoning attacks are a neglected, yet critical, attack scenario. 

In this paper, we propose \sys{}, the first training-time poisoning attack against speech recognition. 
In our design of \sys{}, we focus on \emph{hybrid} \ac{ASR} systems, as they are widely used in practice and for commercial products such as Amazon's Alexa and Sonos's Voice Control~\cite{coucke-20-numberassistants}. 
The goal of our poisoning attack is similar to adversarial example attacks~\cite{carlini2018audio, schonherr2018adversarial, schonherr2020imperio, vaidya2015cocaine}: We want to manipulate such an \ac{ASR} system so that it recognizes potentially problematic commands (e.g., ``open the door``), while the user says something else. 
The difference is that we achieve the desired outcome not by manipulating the \emph{input} utterances to the system, but rather by tampering with its \emph{training data}.

The task of an \ac{ASR} system is to transcribe an audio waveform into a sequence of words. 
For a correct transcription, speech recognition systems consider inherent structures of speech, like the grammar of a language or context dependencies of phonetic units. 
For this purpose, a hybrid system utilizes two models, an \emph{acoustic model} and a \emph{language model}: 
The acoustic model divides an audio waveform into overlapping frames and processes each frame individually, which results in a \emph{sequence} of states, serving a phonetic representation. 
Subsequently, this sequence is decoded with the language model that is trained on linguistic features to predict a transcription. 
From an attacker's perspective, both components and their interplay need to be considered.
Additionally, \ac{ASR} systems are---in general---trained from scratch, and we can therefore not rely on fine-tuning a pre-trained model; a threat model that is often assumed by previous poisoning attacks.

Having considered these challenges, we design and implement \sys{} against hybrid \ac{ASR} systems and evaluate the effectiveness from various aspects that are essential for a realistic attack. 
\sys{} consists of three fundamental steps: First, in the \emph{sequence selection}, we select a target input and define the sequence of target states that corresponds to an attacker-chosen target transcription. 
Since there is no one-to-one mapping between states and the transcription, we perform a frequency analysis on the training data to choose a target sequence that would also occur in natural speech. 
Based on this target sequence, we select poison samples in the training data during the  \emph{poison selection} step. 
Finally, for \emph{poison crafting}, we add malicious perturbations to the raw audio waveform of the selected poison samples. 
To compute such perturbations, we use a set of surrogate models, which are updated at each step of the poison optimization, with the goal that the malicious characteristics of the poisoned data transfer to \emph{any} model trained on the resulting dataset.

To empirically evaluate \sys{}, we perform \emph{single-word} replacement attacks on the TIDIGITS dataset~\cite{leonard1993tidigits}, which is composed of uttered digit sequences of different lengths. 
When poisoning on average only 25.44 seconds of audio (0.17\,\% of the victim's training set), \sys{} achieves attack success rates of more than 83.3\,\%.
We further evaluate \sys{} by performing \textit{multi-word} replacement attacks, where we aim to replace all digits of the target sequence with randomly chosen digits.
To examine the scalability of our approach, we additionally apply \sys{} against the larger \emph{Speech Commands} dataset~\cite{warden2018speech} and show that the attack remains successful. 
For this dataset, having poisoned only 116.73 seconds of audio (0.14\,\% of the training set), \sys{} achieves an attack success rate of 73.3\,\%. 

We verify \sys{}'s practical feasibility and demonstrate that the attack remains viable in over-the-air scenarios by playing the target audio waveforms in both simulated and real~rooms. Furthermore, we study the transferability of the attack and use \sys{}'s poisoned data---generated with a hybrid \ac{ASR} system---to train an \emph{end-to-end} system that is publicly available in the speech toolkit SpeechBrain~\cite{speechbrain} and has an entirely different architecture. For this scenario, we observe an attack transferability rate of~36.4\%.

Finally, we conduct a user study, in which we ask human participants to transcribe the poisoned data.
Such a study has often been missing in prior works, and as noted by Schwarzschild et al.~\cite{schwarzschild2021just}, most current attacks in the visual domain produce easily visible artifacts and distortions.  
For \sys{}, on average, more than 85\% of the poison samples were transcribed into their original labels, showing that \sys{} is able to generate clean-label poison samples. 

\smallskip \noindent
In summary, we make the following key contributions:
\begin{itemize}
    \item \textbf{Poisoning \acs{ASR}.} We propose the first training-time poisoning attack against \ac{ASR} systems and demonstrate that poisoning attacks are a real threat to ASR systems.
    \item \textbf{Full Training.} We assume the victim's system is trained on the poisoned data \emph{from scratch}. As shown by prior work~\cite{schwarzschild2021just}, this is significantly harder than the predominantly studied transfer learning setting.
    \item \textbf{Practical Evaluation.} We consider various aspects that are essential for the deployment of a realistic attack against a speech recognition system. We show that the attack is effective with limited knowledge in over-the-air settings, and that it transfers to unknown \ac{ASR} architectures.
    \item \textbf{Intelligibility.} 
    We conduct a user study and show that the attack generates clean-label poison samples as well as that the original transcription is intelligible. 
    Additionally, we test the effects of psychoacoustics to hide the adversarial noise below the human hearing thresholds.

\end{itemize}

To foster further research in this area, we release the source code of all experiments as well as the poison samples generated by \sys{} at \emph{\url{https://github.com/ucsb-seclab/VenoMave}}.

\section{Technical Background}
\label{sec:background}
The task of an \ac{ASR} system is to automatically transcribe any spoken content from raw audio waveforms into text. Nowadays, these systems can be basically of two kinds: end-to-end systems and hybrid systems. The former refers to neural architectures where the network directly transforms the audio waveform into a character transcription. On the other hand, hybrid DNN/HMM systems combine a neural network with a statistical model; namely, a \ac{DNN} for acoustic modeling and a \acf{HMM}, used as the language model for cross-temporal information integration. 

Compared to end-to-end systems, hybrid systems continue to offer greater flexibility because of their decoupled acoustic and language model. 
This, in turn, makes reusing or fine-tuning the individual models significantly easier and computationally less expensive.
Furthermore, unlike large and monolithic end-to-end systems, the acoustic modeling of hybrid systems can be built closer to the user's personal device and away from the cloud, alleviating the privacy concerns of customers~\cite{coucke-20-numberassistants}.
For these reasons~\cite{wang-19-overview-end2end}, hybrid \ac{ASR} systems continue to be used in practice by commercial products such as Amazon's Alexa, or very recently by Sonos's Voice Control~\cite{coucke-20-numberassistants}. 

Figure~\ref{fig:asr_overview} provides an overview of the main system components of a modern DNN/HMM hybrid system:
\label{sec:prepro}
\begin{itemize}[topsep=3pt, itemsep=3pt, partopsep=3pt, parsep=3pt]
\item \emph{MFCCs Extraction.}
The raw waveform input is typically processed into a feature representation that should ideally preserve all relevant information (\eg phonetic information that describes the smallest acoustic unit of speech) while discarding the unnecessary remainders (\eg acoustic properties of the room). Therefore, the input waveform is divided into overlapping frames of fixed length, and each frame is processed to obtain \acp{MFCC} features~\cite{stevens-37-scale}. \acp{MFCC} features consider the logarithmic frequency perception of the human auditory system and are a very common feature representation for \ac{ASR} systems. 


\item \emph{Acoustic Model DNN.}
At the core of the system, the \ac{DNN} is used as the acoustic model to predict the probabilities for distinct speech sounds (i.e., \emph{phones}) for a given input frame. The phonetic description itself together with context dependencies and language grammar are described by the \ac{HMM} states. Thus, the \ac{DNN} outputs pseudo-posteriors for each input frame, which describe the probabilities for each of the \ac{HMM} states.

\item \emph{Decoder.} 
Given the output matrix of the \ac{DNN}, an optimal path (which is interpreted as a sequence of words) is searched through the \ac{HMM} via dynamic programming (e.g., Viterbi decoding~\cite{omura-69-viterbi}).
\end{itemize}

When training an \ac{ASR} system, the exact alignment between utterances and transcriptions (i.e., the labels) is usually not available. To account for this, \emph{Viterbi training} is commonly utilized.
Starting with training on equally aligned labels, an initial \ac{DNN} is trained, followed by the decoding of the training data, which results in a new and better fitting alignment between utterances and their transcriptions.

\section{Method}
\label{sec:method}

\begin{figure*}[t]
    \centering
    \includegraphics[width=1.6\columnwidth]{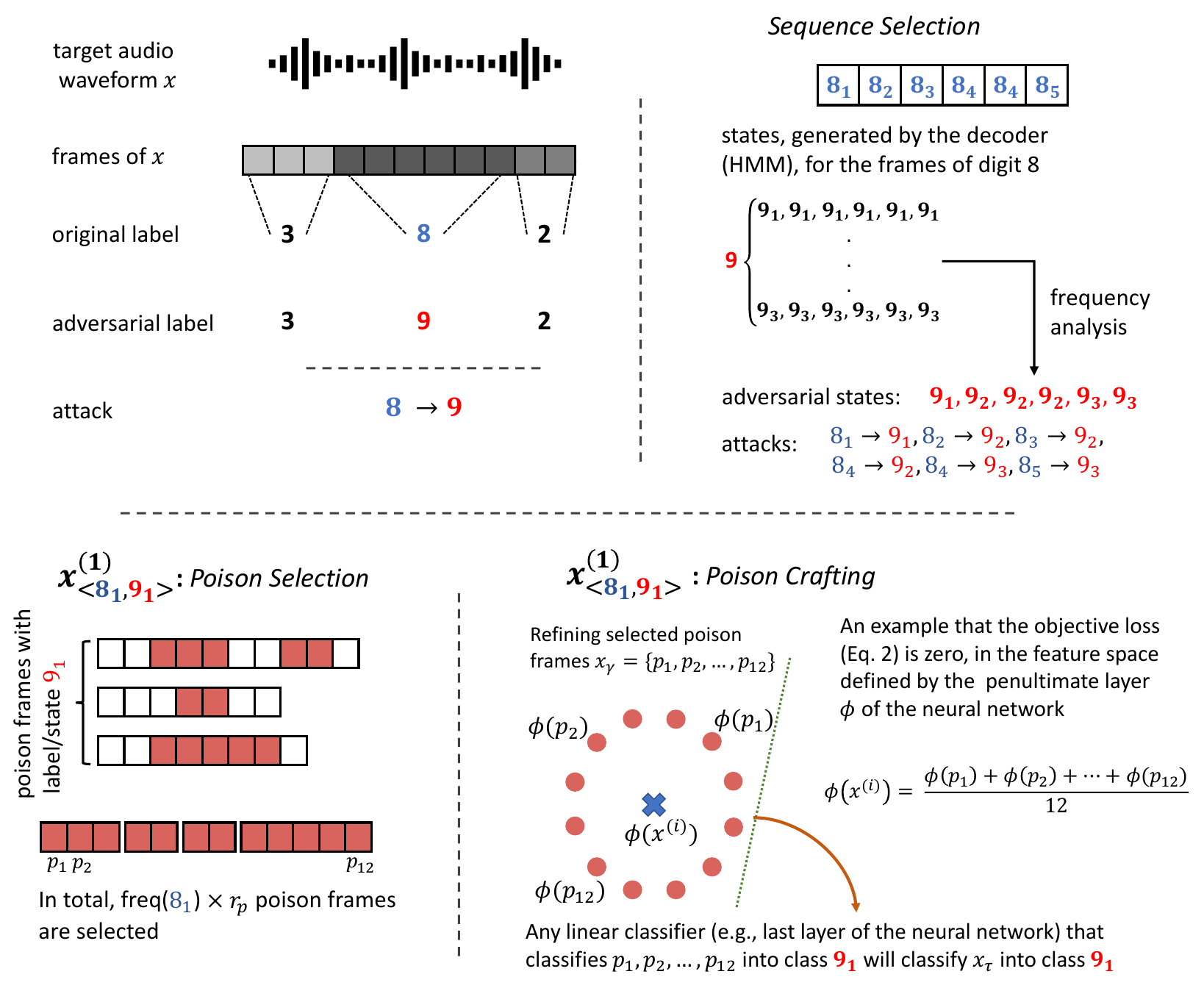}
    \vspace{-0.7em}
    \caption{\textbf{Training-time poisoning attack.} An example of transcribing an utterance with original transcription \texttt{382} into \texttt{392} using \sys{}.
    First, the attacker determines which frames of the audio file need to be targeted and what is the target HMM states of these frames. For each of these frames, an individual poisoning attack is performed to fool the surrogate networks. After a successful attack, the poisons transfer to the victim's network and decode the target transcription \texttt{392}. For simplicity, only the attack for the first frame is depicted, considering only one surrogate model. 
    In practice, an entire time series needs to be attacked successfully.
    } 
    \label{fig:method_eg}
    \vspace{-1em}
\end{figure*}

On a high level, an adversary wants to trigger a targeted misclassification of an unmodified utterance by introducing maliciously altered training samples.
This is a challenging task: 
First, the input of an \ac{ASR} system is a time series and, consequently, the system's output is also a sequence of classes. An adversary needs to consider these time dependencies when crafting poisons.
Second, \ac{ASR} systems are typically trained from scratch, and an attacker needs to take the complete training pipeline into account. This is a much more difficult task compared to the predominately studied poisoning setting of \emph{linear transfer learning}, where only the fine-tuning of a machine learning model is attacked~\cite{schwarzschild2021just}. 

To address these challenges, we introduce \sys{}. In the following, we describe the details of \sys{}'s training-time poisoning attack, starting with the description of our threat~model.

%





\subsection{Threat Model} 
The attacker manipulates data points of the victim's training set, aiming to poison the victim's \ac{ASR} to trigger a \emph{targeted} misclassification of a specific utterance into an attacker-chosen transcription. 
The attacker only modifies fractions of the training data by adding malicious perturbations and cannot manipulate the target utterance itself.
In our threat model, we do not limit the amount of perturbation that we add to poison utterances.
This can potentially cause the poisoned data to have wrong transcription labels.
In Section~\ref{sec:userStudy}, we evaluate the human perception of the poisoned data by conducting a listening transcription test.
%
%

For our experiments, we assume attackers with different levels of knowledge of the victim's training parameters, the architecture of the neural network, and the clean training set. In our most restricted threat model, we assume that the adversary knows neither the victim's training data (except for the injected poisoned data) and training parameters nor the architecture of the neural network.
In this setting, the attacker still uses a dataset with a similar distribution to the victim's dataset.

In any case, we assume that the victim always uses an unknown random seed to train the entire \ac{ASR} system from scratch on the manipulated, poisoned training data.
Finally, to build the language model, we assume that the victim uses a dictionary of phonetic word descriptions that is known to the attacker. This is a legitimate assumption, as there are a few dictionaries that are in wide use and can thus be seen as a quasi-standard for pronunciation models, \eg the CMU pronouncing dictionary for English~\cite{lenzo-14-cmudict}.

%



\subsection{\sys{} Algorithm}
\label{sec:method:overview}

%


For a given target audio waveform, our goal is to create a set of poison samples that replace the original transcription with a target transcription if a model is trained on a dataset that contains the poison data.
At a high level, \sys{} achieves this goal by modifying the selected poisoned utterances to be similar to the target utterance in the feature space of the poisoned model.
Figure~\ref{fig:method_eg} illustrates the individual steps of our attack. For the explanation of \sys{}, we focus on changing exactly one word of the transcription. In this example, the \ac{ASR} system is poisoned to recognize an audio waveform with the original transcription \texttt{382} as \texttt{392}, \ie replacing the original word \texttt{NINE} with the word \texttt{EIGHT}.
We use this example throughout this section to explain each step in detail. The full attack is also described in Algorithm~\ref{alg:method}.

\definecolor{commentcolor}{HTML}{316D94}
\newcommand\mycomment[1]{{\algorithmiccomment{\footnotesize\textcolor{commentcolor}{#1}}}}

\begin{algorithm}
\caption{\sys{}}
\label{alg:method}
\textit{\small Inputs} \\
\hspace*{\algorithmicindent} \(x_t\) \mycomment{Target audio waveform} \\
\hspace*{\algorithmicindent} \(W_t\) \mycomment{Target transcription} \\
\hspace*{\algorithmicindent} \(M\) \mycomment{Number of surrogate models} \\
\hspace*{\algorithmicindent} \(\cleanDataset\) \mycomment{Training dataset}\\\vspace{-1em}
\makeatletter
\newcounter{phase}[algorithm]
\newlength{\phaserulewidth}
\newcommand{\setphaserulewidth}{\setlength{\phaserulewidth}}
\newcommand{\phase}[1]{%
  \vspace{-0.1em}
  \Statex\leavevmode\llap{\rule{\dimexpr\labelwidth+\labelsep}{0em}}\rule{\linewidth}{0.1\phaserulewidth}\vspace{-0.2em}

  \Statex\strut\refstepcounter{phase}\textit{\small Phase~\thephase:~#1}
  \vspace{0.25em}
}
\makeatother

\setphaserulewidth{.7pt}
\begin{algorithmic}[1]

    \phase{Initialization}\vspace{-0.2em}
    \Statex {\footnotesize\textcolor{commentcolor}{We train a reference neural network \(\surrogateDNN{}\) and language model \(\surrogateHMM{}\) on the clean dataset $\cleanDataset$. These are used for poison and sequence selection.}}
    \vspace{0.4em}
    \State \(\surrogateDNN{}\), \(\surrogateHMM{}\) \(\gets\) train($\cleanDataset$) 
    
    \phase{Sequence Selection}\vspace{-0.2em}
    \Statex {\footnotesize\textcolor{commentcolor}{Get the relevant audio frames \(x^{(i)}\) for the target transcription, along with the corresponding \ac{HMM} states \(\{\originalstate_i\}_{i=1}^{\numadvpair}\) with the trained reference models \(\langle\surrogateDNN{}, \surrogateHMM{}\rangle\) (line 2). Perform frequency analysis on \(\cleanDataset\) to select the adversarial sequence (line 3).}}
    \vspace{0.4em}
    \State \(x^{(i)}\), \(\{\originalstate_i\}_{i=1}^{\numadvpair} \gets\) get\_target\_frames(\(\langle\surrogateDNN{}, \surrogateHMM{}\rangle\), \(x_t\)) \vspace{0.2em}
    
    \State \(\{\adversarialstate_i\}_{i=1}^{\numadvpair} \gets\) select\_adv\_states(\(\surrogateHMM{}\), \(\cleanDataset\), \(W_t\)) \vspace{0.2em}


    \phase{Poison Selection}\vspace{-0.2em}
    \Statex {\footnotesize\textcolor{commentcolor}{For each attack pair \(T = \{\audioframe^{(\indexone)}_{<\originalstate_\indexone, \adversarialstate_\indexone>}\}_{\indexone = 1}^{\numadvpair}\)  select poison frames \(\poisonDataset{}_i\).}}
    \vspace{0.4em}

    \For{$i=1$ {\bfseries to} \(\numadvpair\)}
        \State \(\poisonDataset{}_i \gets \) select\_poison\_frames(\(\cleanDataset\), \(\originalstate_i\), \(\adversarialstate_i\)) 
    \EndFor
    
    \phase{Poison Crafting}\vspace{-0.2em}
    \Statex {\footnotesize\textcolor{commentcolor}{In each round~k, we retrain surrogates from scratch on the current (poisoned) dataset \(\mathcal{D}\) (lines 9-11). We iteratively update poisons with respect to \(\nabla \textrm{loss}\) (lines 12-19) calculated via Equation~\eqref{eq:BPLoss} and subsequently update $\mathcal{D}$ (line 20). After each round~k, we test \(\mathcal{D}\) with a (surrogate) victim model \(\surrogateDNN{}_\mathcal{V}\) (line 22).}}
    \vspace{0.4em}
    \State \(\mathcal{D} \gets \cleanDataset{}\)
    
    \For{k \(=1\) {\bfseries to} \(\maxiter\)}
    
    \For{$m=1$ {\bfseries to} \(M\)}
        \State \(\surrogateDNN{}_m\), \(\surrogateHMM{}_m\) \(\gets\) train(\(\mathcal{D}\)) 
    \EndFor
    
    \While{not converged}
        \State loss \(\gets\) 0
        \For{\((x^{(i)}, Y_i, Z_i) \gets \advpairsset\)}
        \State loss \(\gets\) loss + $\mathcal{L}$(\(x^{(i)}\), \(\poisonDataset_i\), \(\{\surrogateDNN{}_m\}_{m=1}^{m}\)) 
        \EndFor
        \State loss \(\gets\) \(\frac{\textrm{loss}}{\numadvpair}\) 
        \State update \(\{\poisonDataset_i\}_{i=1}^{\numadvpair}\) using \(\nabla \textrm{loss}\)
    \EndWhile
    
    \State \(\mathcal{D} \gets \)update\_dataset(\(\cleanDataset{}, \{\poisonDataset_i\}_{i=1}^{\numadvpair}\)) 
        
    \State \(\surrogateDNN{}_\mathcal{V}\), \(\surrogateHMM{}_\mathcal{V}\) \(\gets\) train(\(\mathcal{D}\))
    \State \textbf{break} if attack is successful (early stopping) 
    
    \EndFor

\end{algorithmic}
\end{algorithm}

\label{sec:poisonselection}


Considering the hybrid speech recognition architecture, we have to inject poison samples such that the trained acoustic model generates an output sequence that will be decoded as the target words by the language model. 
Therefore, the adversarial label for the acoustic model is a sequence of \ac{HMM} states that describes our target transcription.
Note that not only one possible sequence of states would lead to a specific transcription, as a large number of state sequences map to the same transcription.
For this reason, we first have to determine which state sequence is a promising candidate to achieve the desired output transcript.

To choose the sequence as well as select candidate samples to poison, \sys{} relies on a reference \ac{ASR} system, which is trained on the clean training set.
We refer to this system as \(\langle\surrogateDNN{}, \surrogateHMM{}\rangle\), where \(\surrogateDNN{}\) and \(\surrogateHMM{}\) denote the acoustic model and language model, respectively.

\subsubsection{Sequence Selection} The language model \(\surrogateHMM{}\) defines the word~$\word$ as a sequence of states $\word\!\!=\!\![\hmmstate_\wordstateindex]$ with $\wordstateindex = 1,\dots,\mathcal{K}$.
Assuming that the sequences for the digits \texttt{EIGHT} and \texttt{NINE} consist of 5 and 3 states, respectively, the two words can be described with \ac{HMM} states $\text{\textbf{EIGHT}}\!\!=\!\![8_1, 8_2, 8_3, 8_4, 8_5]$ and $\text{\textbf{NINE}}\!\!=\!\![9_1, 9_2, 9_3]$.
In general, the number of frames of an uttered word is larger than the number of \ac{HMM} states. 
That is, for the word \texttt{NINE} uttered across 6 frames, both sequences \([9_1, 9_1, 9_2, 9_2, 9_3, 9_3]\) and \([9_1, 9_1, 9_1, 9_2, 9_2, 9_3]\) could be selected as the target. 
However, a sequence should be selected that is more probable to be decoded as \texttt{NINE}.
Hence, we look at the appearances of the word \texttt{NINE} in the dataset and select the most common pattern as our target sequence. 

Using \(\langle\surrogateDNN{},\surrogateHMM{}\rangle\), we calculate the relative frequency of state~\(\hmmstate_\kappa\) as the average number of its occurrences in utterances of \texttt{NINE}. Then we select a target sequence that has a distribution of relative frequencies similar to what we have observed in the dataset.
%
Therefore, in our running example, the original sequence \([8_1, 8_2, 8_3, 8_4, 8_4, 8_5]\) should be changed to \([9_1, 9_2, 9_2, 9_2, 9_3, 9_3]\), as the state $9_2$ appears three times more often in the training set than the state $9_1$.
%
%
We then divide our attack into \(N\!=\!6\) smaller poisoning attacks, described by a set~$\advpairsset\!\!=\!\! \{\audioframe^{(\indexone)}_{<\originalstate_\indexone, \adversarialstate_\indexone>}\}_{\indexone = 1}^{\numadvpair}$ of frames~$\audioframe^{(\indexone)}_{<\originalstate_\indexone, \adversarialstate_\indexone>}$ with an original state ~$\originalstate_\indexone$ and an adversarial state~$\adversarialstate_\indexone$. In our example in Figure~\ref{fig:method_eg} the poisoning set is described as

\vspace{-.7em}
\begin{align}
\Big\{\audioframe^{(1)}_{<8_1, 9_1>}, \audioframe^{(2)}_{<8_2, 9_2>}, \audioframe^{(3)}_{<8_3, 9_2>}, \audioframe^{(4)}_{<8_4, 9_2>}, \audioframe^{(5)}_{<8_4, 9_3>},\audioframe^{(6)}_{<8_5, 9_3>}\Big\}\notag.
\end{align}

\subsubsection{Poison Selection} 
We select poison utterances in training data based on the chosen target sequence:
For each attack pair $\audioframe^{(i)}_{<\originalstate_i, \adversarialstate_i>}$, we select poison frames \(\poisonDataset_i\) with label~$\adversarialstate_i$ from one or more utterances.
We use the frequency of the original state \(\originalstate_i\) to determine the number of poison frames to be 
\begin{align}
    \big\lceil\textrm{freq}(\hmmstate\!\!=\!\!\originalstate_i) \cdot \poisonbudget\big\rceil,
    \label{eq:freq}
\end{align}
where \(0\!<\!\poisonbudget\!<\!1\) describes the \emph{poison budget}.
Thus, if an original state~$\originalstate_\indexone$ occurs twice as often in the training set as another original state~$\originalstate_\indextwo$, we also select twice as many poison frames for the attack~\(\audioframe^{(\indexone)}_{<\originalstate_\indexone, \adversarialstate_\indexone>}\) than for the attack~\(\audioframe^{(\indextwo)}_{<\originalstate_\indextwo, \adversarialstate_\indextwo>}\).
The intuition behind this choice is that the attack might fail if the target frame \(x^{(i)}\) has adjacent neighbor frames from its class $\originalstate_i$ in the victim's training set. This has also been observed in prior work~\cite{zhu2019transferable}.
The poison frames---no matter how well they are crafted---need to compete with these neighbor frames to successfully inject the malicious decision boundaries during the training phase.

Our attack only perturbs particular frames of selected poisoned audio files.
This allows to distribute poison frames over multiple utterances, with each utterance consisting of mostly clean frames and only a few poison frames.

\subsubsection{Poison Crafting}
\label{sec:poisoncrafting}
The goal of this step is to modify the selected poison utterances such that they are ``close enough'' to the target utterance in the feature spaces of the surrogate poisoned models after being trained on the poisoned dataset.
The motivation behind this goal is the mathematical guarantee that any \textit{linear} classifier that associates a set of samples~\(\numpois\) to class~$\adversarialstate$ will also classify any point inside their convex hull as class~$\adversarialstate$. 
Specifically, we divide the network into two parts: (1) all layers up to the penultimate layer, named the feature\footnote{Throughout the paper, by the term \emph{features} we refer to the features represented by the penultimate layer, not MFCCs.} extractor network \(\model\), and (2) the last layer, which is a linear classifier.
The victim's model will identify the target frame \(x^{(i)}\) as the target class $\adversarialstate_i$ if $\model(x^{(i)})$ lies within the convex hull of class $\adversarialstate$ formed by the poison frames~\(\{\model(\poisonframe^{(\poisonindex)})\}_{\poisonindex=1}^\numpois\).

For each attack pair $\audioframe^{(i)}_{<\originalstate_i, \adversarialstate_i>}$, we use \(\nummodels\) surrogate models (i.e., similar models trained with different seeds) to optimize the poison frames~\(\poisonDataset_i=\{\poisonframe^{(\poisonindex)}\}_{\poisonindex=1}^\numpois\) with the following loss: 

\small
\vspace{-1em}
\begin{align}
\mathcal{L} := \underset{\{\poisonframe^{(\poisonindex)}\}}{\textrm{min}} \; &\frac{1}{2\nummodels} \sum_{\modelindex=1}^\nummodels \frac{\norm{\model^{(\modelindex)}(x^{(i)}) - \frac{1}{\numpois}\sum_{\poisonindex=1}^{\numpois} \model^{(\modelindex)}(\poisonframe^{(\poisonindex)})}^2}{\norm{\model^{(\modelindex)}(x^{(i)})}^2}
\label{eq:BPLoss}
\end{align}
\normalsize

To solve this non-convex problem, we iteratively apply gradient descent to optimize the poison frames~\(\poisonDataset_i\).

Our motivation behind optimizing Equation~\ref{eq:BPLoss} over \(\nummodels\) surrogate models is based on prior work~\cite{zhu2019transferable, aghakhani2020bullseye} that relies on the assumption that by obtaining the above heuristics for similar models, such a guarantee will also transfer to unknown victim models. 
These attacks presented high success rates against \emph{linear transfer learning}, where a pre-trained but \emph{frozen} network \(\model\) is used to calculate features for an application-specific linear classifier, which is fine-tuned on the poisoned dataset.
However, as shown by Schwarzschild et al.~\cite{schwarzschild2021just}, such heuristics will not hold when the victim's model is trained on the poisoned dataset from scratch, as the feature space is also altered during training.
In fact, we made similar observations in preliminary experiments.

To cope with this challenge, we train a set of surrogate networks \(\{\surrogateDNN{}_m\}_{m=1}^{M}\) \emph{from scratch} on the current (poisoned) dataset at the beginning of each round of the attack.
Subsequently, we modify the poison samples to achieve our desired heuristics with respect to the refreshed surrogate models.
Our intuition is that after several rounds of the attack we reach a state in which the poisoned data needs no further modifications to obtain the heuristics. 
To check whether this happens or not, at the end of each round of the attack, we train a (surrogate) victim \ac{ASR} system on the current poisoned dataset from scratch.
The attack terminates if either it succeeds against this \ac{ASR} system (early stop) or we reach a maximum number of rounds \(\maxiter\).

For the evaluation of \sys{}, we consider an attack to be successful if and only if it succeeds against the target victim's \ac{ASR} system, where both the neural network and language model components are trained on the poisoned dataset from scratch.
Our experiments demonstrate that the malicious characteristics of our crafted poisoned data successfully transfer to the victim's poisoned~model with high probability. 

\section{Evaluation}
\label{sec:exp}

In this section, we empirically assess \sys{} in a series of experiments.
We start by evaluating the attack's efficacy on the task of recognizing sequences of digits with the TIDIGITS dataset~\cite{leonard1993tidigits}. Building upon this, we consider a larger \ac{ASR} system that is trained on the \emph{Speech Commands} dataset~\cite{warden2018speech}.
Our experiments show that the attack is effective in poisoning \ac{ASR} systems, remains viable with limited knowledge about the victim's system and in over-the-air settings.
Furthermore, we demonstrate that the malicious characteristics of the poisoned data---crafted with \sys{} for a hybrid \ac{ASR} system---transfer to an \emph{end-to-end} system.
%
Throughout the experiments, we use the open-source \ac{ASR} system used by Däubener et al.~\cite{daeubener-20-uncertainty} for studying evasion attacks against \ac{ASR} systems.

\subsection{Metrics}

Before we get into the details of our results, we describe the standard measures used to assess the quality of the poison samples, both in terms of effectiveness as well as conspicuousness.

\subsubsection{Attack Success Rate}
In all experiments, an attacker aims to induce a targeted misclassification for a single utterance. 
If the targeted misclassification is not triggered, we consider the attack as failed. 
The \emph{attack success rate} then describes the percentage of successful~attacks.



\subsubsection{Clean Test Accuracy}
We evaluate the victim's performance against the test set to calculate the \textit{clean test accuracy} of the model.
An ideal poisoning attack does not degrade the model performance for non-target inputs; otherwise, it might be suspicious.
For all test samples, given the model transcriptions, we count and accumulate all substituted words~\(\substitutions\), inserted words~\(\insertions\), and deleted words~\(\deletions\) to calculate the accuracy via 
\begin{align}
    \text{accuracy} = \frac{\numwords - \insertions - \substitutions - \deletions}{\numwords}\notag,
\end{align}
where \(\numwords\) is the total number of words in the test set's ground-truth labels.

\subsubsection{Segmental Signal-to-Noise Ratio (SNRseg)}
To quantify the magnitude of required changes, we use the Segmental Signal-to-Noise Ratio (SNRseg).
This metric measures the amount of noise~$\noise$ added by an attacker to the original signal~\(\inputaudio\) and is computed~via
%
\begin{align}
\text{SNRseg(dB)} = \dfrac{10}{\numframes} \sum_{\frameindex=0}^{\numframes-1} \log _{10}  \frac{\sum_{\timeframesindex=\numtimeframes\frameindex}^{\numtimeframes\frameindex+\numtimeframes-1} \inputaudio^2(\timeframesindex)}{\sum_{\timeframesindex=\numtimeframes\frameindex}^{\numtimeframes\frameindex+\numtimeframes-1} \noise^2 (\timeframesindex)}\notag,
\end{align}
%
where $T$ is the segment length and $K$ the number of segments. Thus, the higher the SNRseg, the \emph{less} noise has been added. We use a frame length of $12.5$\,ms, which corresponds to $T=200$ at a sampling frequency of $16$\,kHz.
As only very small parts of the poison files are changed, we measure the SNRseg only for the poisoned~frame (i.e., clean parts of the poison samples are excluded) to provide a fair assessment of the added noise. 

\begin{table}[t]
\addtolength{\tabcolsep}{-4pt}
\caption{Neural network architectures used in experiments. 
Networks use two or three hidden layers, each with a softmax output layer of size 95, corresponding to the number of \ac{HMM} states. 
The baseline test accuracy is for when the victim uses a clean dataset.}

\newcolumntype{C}[1]{>{\centering\let\newline\\\arraybackslash\hspace{0pt}}m{#1}}

\centering
\smallskip
\small
\begin{tabular}{l|l C{4cm} C{4cm}} 
\toprule 
\textbf{Name} & \textbf{Layer description} & \textbf{\# Parameters} \\ 
\midrule
\netTwoLayerBold{}                      & \((100, 100)\) neurons         & ~54,895\\ 
\netTwoLayerPlusBold{} 	                & \((100, 200)\)	neurons         & 100,095\\ 
\textbf{\netThreeLayerBold{}} 		    & \((100, 100, 100)\) neurons 	& ~64,995\\ 
\textbf{\netThreeLayerPlusBold{}} 		& \((400, 300, 200)\) neurons	    & 340,395\\ 

\bottomrule
\end{tabular}
\label{table:nets}
\vspace{-1em}
\end{table}

\subsection{ Attack Parameters}
\label{exp:tidigit}

We first evaluate the attack efficacy with respect to its salient parameters: the number of surrogate models as well as varying sizes of the poison budget. 
For this experiment, we consider a threat model, where the attacker has full knowledge of the victim's network architecture, training parameters, and training set.
The adversary uses this knowledge to train surrogate \ac{ASR} systems for poison optimization.
We run each attack instance for a maximum of \(\maxiter=20\) rounds. For the early stopping criteria, we test after each round if we succeed against a (surrogate) test model. 


\subsubsection{Experimental Setup}
We use the TIDIGITS dataset~\cite{leonard1993tidigits}, which is designed for speaker-independent recognition of digit sequences and consists of eleven words: \texttt{ONE}, \texttt{TWO}, ..., \texttt{NINE}, \texttt{ZERO}, and \texttt{OH}. We use 8,623 utterances for the training set and 4,390 utterances for the test set. The sequences are spoken by 225 speakers (111 men and 114 women), which are split equally into disjoint sets between the training and test set.
For our poisoning attack trials, we randomly sample 30 single-digit utterances among the 4,390 test samples and assign a target label to each of them. Target labels are chosen randomly and are different from the ground-truth transcription. 

The victim's \ac{ASR} system uses the \netTwoLayerPlus{} architecture (described in Table~\ref{table:nets}) with a softmax output layer of size 95, corresponding to the number of \ac{HMM} states. This system is trained from scratch for 33 epochs with a batch size of 32 using the Adam~\cite{kingma2014adam} optimizer with a learning rate of $1e^{-4}$. This training also includes three epochs of Viterbi training to build the language model. Hyperparameters were chosen to maximize the clean test accuracy. For the baseline model ---only trained with clean data--- we achieved a test accuracy of $98.79$\,\%.

For evaluation of the attack, the random seed used by the victim is unknown. Thus, the specific parameters of the victim's \ac{ASR} system, the neural network, and the \ac{HMM}---which depend on the neural network due to Viterbi training---are not used during poison optimization.

To accelerate the attack, we freeze the \ac{HMM} component and only train the \ac{DNN} for the surrogate \ac{ASR} systems.
We found this effective as the language model does typically not change significantly.
The frozen surrogate \ac{HMM} is trained in advance by training an \ac{ASR} system for $15$ epochs on the clean training set, followed by three epochs of Viterbi training.
During the attack, we train the surrogate \ac{ASR} systems for 25 epochs until convergence. 

\subsubsection{Results}
\begin{table}[t]
\addtolength{\tabcolsep}{-3pt}
\vspace{-.8em}
\caption{Evaluation of \sys{} when it uses different numbers of surrogate networks. 
The \(\poisonbudget\) is set to 0.005. 
This experiment was performed on a machine with NVIDIA RTX A6000 graphics cards (with CUDA 11.0, PyTorch 1.9.1, and Torchaudio 0.9.1). Note that as \sys{} employs an early-stopping procedure (see Algorithm~\ref{alg:method}), increasing \(\mathbf{\nummodels}\) will not necessarily lead to a longer attack time.
}
\vspace{-0.6em}
\centering
\smallskip
\small
\begin{tabular}{l|r r r r r r} 
\toprule
 & \multicolumn{6}{c}{\(\mathbf{\nummodels}\)} \\ \cmidrule[0.6pt]{2-7}
 & \multicolumn{1}{c}{\textbf{1}} & \multicolumn{1}{c}{\textbf{2}} & \multicolumn{1}{c}{\textbf{4}} & \multicolumn{1}{c}{\textbf{6}} & \multicolumn{1}{c}{\textbf{8}} & \multicolumn{1}{c}{\textbf{10}} \\
\toprule
\textbf{\# Attack step (\(\bm{\maxiter}\))}     & 15.7         & 11.5             & 7.9           & 7.6       & 6.8      & 7.0         \\
\textbf{Attack time (hours)}                 & 1.54          & 1.36              & 1.46          & 3.43      & 3.33      & 5.33        \\
\midrule
\textbf{Clean test acc. (\%)} 		        & 97.84         & 97.84             & 97.81         & 97.79     & 97.84     & 97.81       \\
\midrule
\textbf{Attack succ. rate (\%)}	            & 43.3         & 76.7             & 80.0         & 80.0     & 86.7     & 83.3       \\

\bottomrule
\end{tabular}
\label{table:exp:diffnumsurroagtenetworks}
\end{table}
We first evaluate the attack success rate as a function of the number of surrogate models.
Table~\ref{table:exp:diffnumsurroagtenetworks} presents the performance of \sys{} for different numbers of surrogate networks.
Note that a higher number of surrogate models adds to the complexity of Equation~\ref{eq:BPLoss}. However, more surrogate networks can help the attack to succeed in fewer steps and, consequently, this increased complexity does not necessarily lead to a longer attack time. This is also evident from the results in Table~\ref{table:exp:diffnumsurroagtenetworks}.
We obtain the highest attack success rate (86.7\,\%) for \(\nummodels=8\) surrogate models. 
In the case where we use \(\nummodels=10\) surrogate models, the attack time and required attack steps are increased while a lower attack success rate is obtained. Note that the number of attack steps K in Table~\ref{table:exp:diffnumsurroagtenetworks} is the average number for all 30 poisoning trials for each entry.


\begin{table}[t]
\caption{Evaluation of \sys{} when the poison budget \(\poisonbudget\) is successively increased from 0.001 to 0.01. 
} 
\vspace{-0.6em}
\addtolength{\tabcolsep}{-3pt}
\centering
\smallskip
\small
\begin{tabular}{l|r r r r} 
\toprule
 & \multicolumn{4}{c}{\(\mathbf{\poisonbudget}\)} \\ \cmidrule[0.6pt]{2-5}
 & \multicolumn{1}{c}{\textbf{0.001}} & \multicolumn{1}{c}{\textbf{0.003}} & \multicolumn{1}{c}{\textbf{0.005}} & \multicolumn{1}{c}{\textbf{0.01}} \\
\midrule
\textbf{Poison data length (seconds)}      & 6.20                & 15.93               & 25.44               & 48.73                       \\
\textbf{\# Poison data samples}               & 96.23               & 248.10               & 387.83             & 693.57                    \\
\midrule
\textbf{Clean test accuracy (\%)} 		    & 97.85             & 97.84             & 97.84            & 97.76         \\
\midrule
\textbf{Attack success rate (\%)}	            & 23.3             & 76.7             & 86.7    & 83.3                     \\
\bottomrule
\end{tabular}
\label{table:exp:diffpoisonbudget}
\vspace{-1em}
\end{table}

Next, we evaluate \sys{} for varying levels of poison budget~\(\poisonbudget\) (see Section~\ref{sec:poisoncrafting}). The results are shown in Table~\ref{table:exp:diffpoisonbudget}.
We observe a general trend that an increase of the poison budget leads to a higher attack success rate (23.3\,\% $\rightarrow$ 83.3\,\%), which stagnates for poison budgets larger than $0.005$. A higher budget allows the attacker to manipulate an increasing number of poison frames and, thus, has more control over the training process. However, from a certain number, this effect is less distinct as the surrogate models also need to maintain a good clean test accuracy.
The general improvement comes at a price; the length and number of the poisoned data increases (6.20\,s $\rightarrow$ 48.73\,s) from a total of 15,254\,s training data. 
We observe the best performance with a budget \(\poisonbudget\!=\!0.005\), where we poison only 0.17\,\% of the training set while achieving an attack success rate of~86.7\,\%.

\begin{figure*}[t]
    \vspace{-1.5em}
  \centering
  \begin{subfigure}{0.49\textwidth}
      \centering
      \includegraphics[trim=0 5 0 0, clip, width=\textwidth]{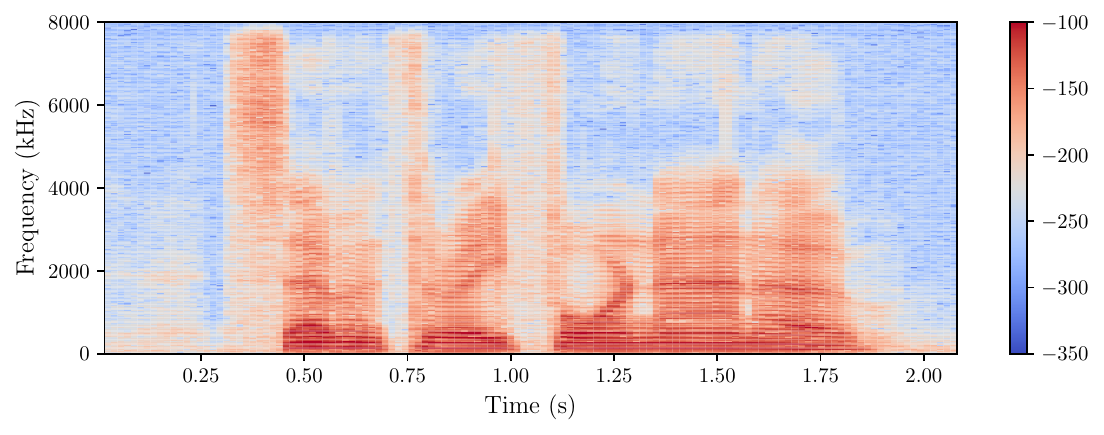}
      \vspace{-1.5em}
      \caption{Original Signal}
      \label{fig:spectogram-referenceA}
  \end{subfigure}
  \hfill
  \begin{subfigure}{0.49\textwidth}
      \centering
      \includegraphics[trim=0 5 0 0, clip, width=\textwidth]{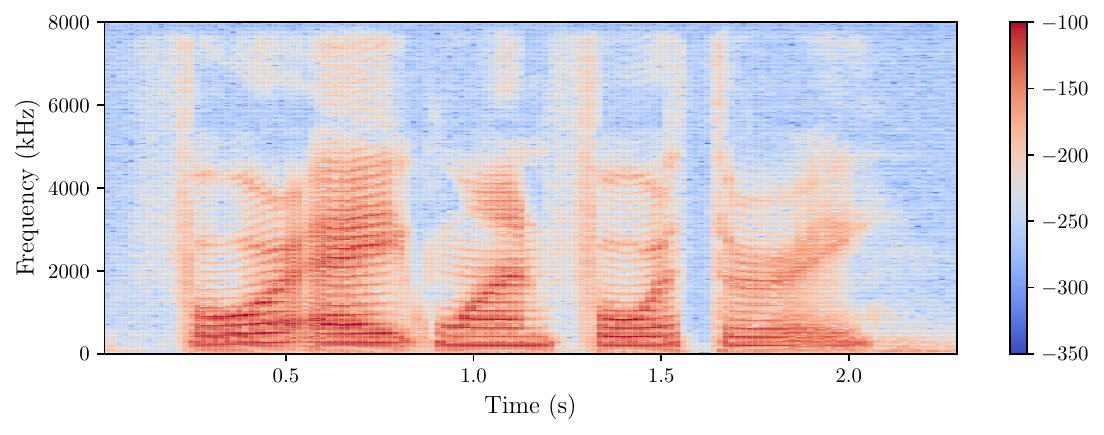}
      \vspace{-1.5em}
      \caption{Original Signal}
      \label{fig:spectogram-referenceB}
  \end{subfigure}
  \begin{subfigure}{0.49\textwidth}
      \centering
      \includegraphics[trim=0 20 0 0, clip, width=\textwidth]{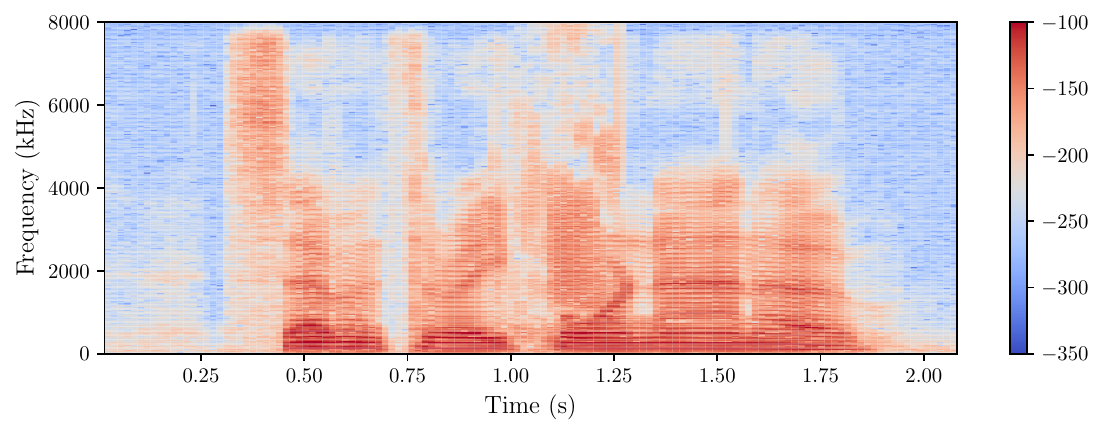}
      \vspace{-1.5em}
      \caption{Poison Signal}
      \label{fig:spectogram-poisonA}
  \end{subfigure}
  \hfill
  \begin{subfigure}{0.49\textwidth}
      \centering
      \includegraphics[trim=0 20 0 0, clip, width=\textwidth]{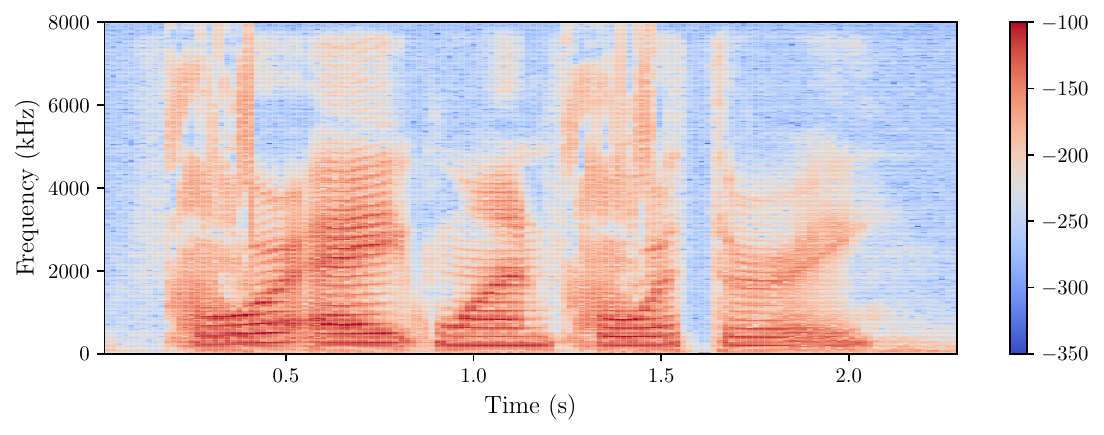}
      \vspace{-1.5em}
      \caption{Poison Signal}
      \label{fig:spectogram-poisonB}
      \end{subfigure}
  \begin{subfigure}{0.49\textwidth}
      \centering
      \includegraphics[trim=0 20 0 0, clip, width=\textwidth]{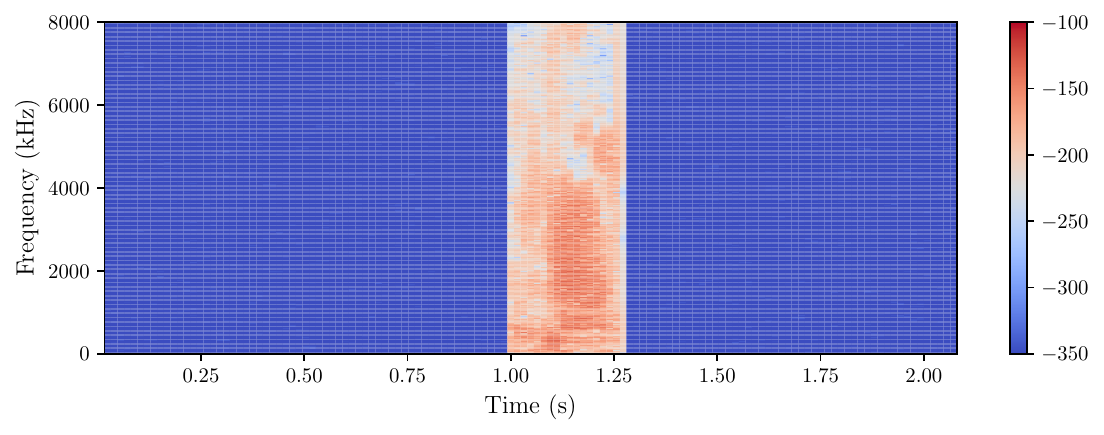}
      \vspace{-1.5em}
      \caption{Difference}
      \label{fig:diffA}
  \end{subfigure}
  \hfill
  \begin{subfigure}{0.49\textwidth}
      \centering
      \includegraphics[trim=0 20 0 0, clip, width=\textwidth]{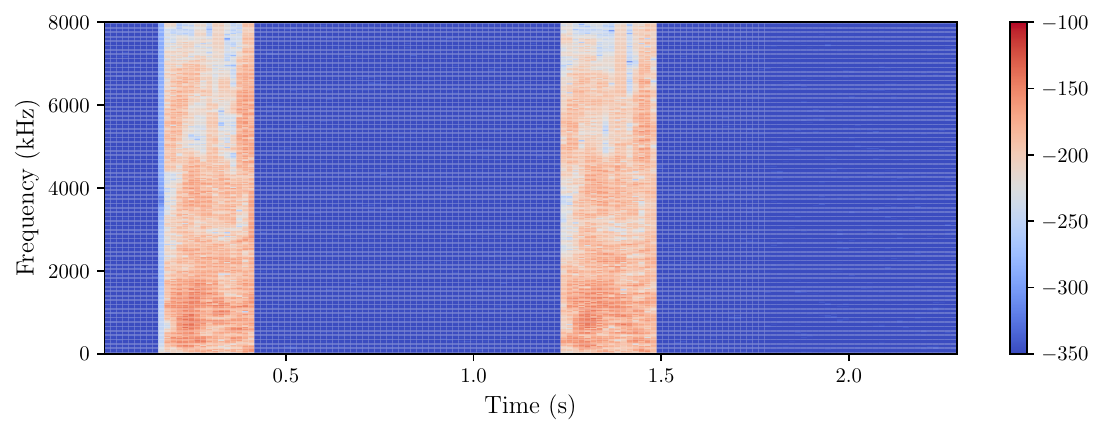}
      \vspace{-1.5em}
      \caption{Difference}
      \label{fig:diffB}
  \end{subfigure}
  \vspace{-0.4em}
    \caption{\textbf{Spectrograms of Poisons.} We present two example poisons computed with \sys{}. The left column shows an utterance of digit sequence \texttt{SEVEN, THREE, FOUR, NINE, OH} and the right shows an utterance of digit sequence \texttt{FOUR, EIGHT, ONE, FOUR, THREE}. Both poison the digit \texttt{FOUR} to \texttt{OH}. Figure~\ref{fig:spectogram-referenceA} and \ref{fig:spectogram-referenceB} show the unmodified signals, Figure~\ref{fig:spectogram-poisonA} and \ref{fig:spectogram-poisonB} depict the poison version, and Figure~\ref{fig:diffA} and \ref{fig:diffB} show the respective differences of both versions.
    }
  \vspace{-1.5em}
  \label{fig:spectogram}
\end{figure*}

Figure~\ref{fig:spectogram} shows an example of a poisoned audio file as well as its respective original audio file.



\begin{table}[t]
\vspace{-0.8em}
\caption{The attack performance for unknown training parameters and network architectures.}
\vspace{-0.6em}

\centering
\smallskip
\small
\label{table:diffVictimSettingsSmallTable}
\begin{tabular}{l|r r r} 
\toprule
 & \multicolumn{3}{c}{\textbf{Victim's network}} \\ \cmidrule[0.6pt]{2-4}
 & \multicolumn{1}{c}{\netTwoLayerBold{}} & \multicolumn{1}{c}{\netThreeLayerBold{}} & \multicolumn{1}{c}{\netThreeLayerPlusBold{}} \\
 \midrule
\textbf{Baseline test accuracy (\%)}                           & 98.75             & 98.41      & 99.01 \\ 
\textbf{Clean test accuracy (\%)} 		                 & 97.92             & 98.04     & 99.02         \\
\textbf{Attack success rate (\%)}	                     & 86.7             & 86.7     & 83.3                     \\

\bottomrule
\end{tabular}
\end{table}
\begin{table}[t]
\centering
\small
\addtolength{\tabcolsep}{-3.5pt}
\caption{Evaluation of \sys{} for partial and unknown set of clean training samples. The victim uses different training parameters than the attacker. We divide the training set of TIDIGITS into two subsets, with ``Split 1'' containing the first half and ``Split 2'' containing the second half of the speakers (56 speakers each).}
\vspace{-0.4em}
\label{table:exp:diffVictimTrainingSets}
\begin{tabular}{ll|ll|cc}
\toprule
\multicolumn{2}{c|}{\textbf{Attacker}} & \multicolumn{2}{c|}{\textbf{Victim}} & \textbf{Clean test}  & \textbf{Attack succ.}\\
\multicolumn{1}{c}{\textbf{Network}} & \multicolumn{1}{c|}{\textbf{Tr. set}} & \multicolumn{1}{c}{\textbf{Network}} & \multicolumn{1}{c|}{\textbf{Tr. set}} & \textbf{acc. (\%)} & \multicolumn{1}{c}{\textbf{rate (\%)}} \\
\midrule
\multirow{2}{*}{\netTwoLayerPlus{}} & \multirow{2}{*}{Split 1} & \multicolumn{1}{l}{\netThreeLayer{}} & Split 2 & \multicolumn{1}{c}{97.92} & \multicolumn{1}{c}{86.7} \\ 
 &  & \multicolumn{1}{l}{\netThreeLayer{}} & Split 1 + 2 & \multicolumn{1}{c}{98.03} & \multicolumn{1}{c}{80.0} \\
 \bottomrule
\end{tabular}
\vspace{-1em}
\end{table}

\subsection{Limited-Knowledge Adversary}
\label{exp:limitedKnowledgeAdversary}

For most applications in practice, it is unrealistic to assume that an adversary has detailed knowledge of the exact training parameters, architecture, and the training data that is used by the victim. In the following, we therefore want to relax the threat model and consider an adversary with limited knowledge. We consider two settings: (1) First, we restrict access to the victim's model architecture and training parameters, and (2) second, we extend the knowledge limitations and additionally restrict access to the victim's training data (except for the poisoned data).
For both settings and based on the previous experiments, we set the poison budget to \(\poisonbudget=0.005\) and consider \(\nummodels=8\) surrogate models.

\subsubsection{Model Architecture and Parameters}
%
We consider that the victim uses one of three different model architectures: \netTwoLayer{}, \netThreeLayer{}, or \netThreeLayerPlus{} from Table~\ref{table:nets}. All models are trained from scratch for 32 epochs, of which epochs 11 and 12 include Viterbi training. The victim uses Adam with a learning rate of \(4e^{-4}\), a batch size of 64, and a dropout probability of 0.2. The dropout layer is added after the first hidden layer. 


Table~\ref{table:diffVictimSettingsSmallTable} shows that the malicious characteristics of the poisoned data remain even if the victim uses different training parameters and network architectures. Also, for all models the clean test accuracy remains almost the same in comparison to the baseline test accuracy, which measures the accuracy of the models trained on exclusively clean data.
It is worth noting that in prior work, dropout was typically disabled, as in a transfer learning scenario, a rational victim will usually overfit the training set~\cite{aghakhani2020bullseye, zhu2019transferable}.
Since this is usually not the case when the victim's model is trained from scratch, we enable dropout in this experiment.
Our results show that the poisoned data survive the randomness introduced by the dropout.


\subsubsection{Training Dataset} 
\label{exp:limitedKnowledgeOfTrainingSet}

Building upon the previous experiment, we further reduce the attacker's knowledge and assume that the attacker only has partial knowledge about the training set of the victim and its underlying distribution.
In general, the adversary uses their knowledge about the training data to (1) perform the ratio analysis (see Section~\ref{sec:method}) and (2) train surrogate networks for the poison crafting step. Note that for this experiment we continue to use an unknown victim's model architecture.

For the experiment, we divide the training data into two subsets with disjoint sets of 56 speakers each.
We restrict the adversary to access only the first subset (Split 1, 56 speakers). For the victim, we consider two different scenarios: (1) training samples only from the second subset (Split 2, 0\,\% overlap), and (2) the entire training set (Split 1+2, 50\,\% overlap). Similar to the previous experiment, we evaluate a victim with different training parameters and network architecture (\netThreeLayer{}).
As the poison samples only depend on Split 1, we use the same data for both cases. 


Table~\ref{table:exp:diffVictimTrainingSets} presents the performance of \sys{} for these two scenarios. 
When the victim's training set has no overlap with the attacker's training set, \sys{} achieves an attack success rate of 86.7\,\%.
When the attacker's training set consists of 50\,\% of the victim's training set, \sys{} achieves an attack success rate of~80\,\%.
While the same poisoned data is used in these two cases, in the latter case, the poisoned data are competing with more clean data points.
This may explain why \sys{} achieves a lower attack success rate despite the fact that it has partial knowledge of the victim's training set.
The average clean test accuracy is 97.92\% and 98.03\% for 0\,\% and 50\,\% overlap cases, respectively.

\subsection{Multi-Word Replacement Attack}
\label{exp:multiword}
Next, we want to scale the attack to more complex targets and, in particular, aim to replace multiple words. 
This can be realized by launching multiple individual word replacement attacks simultaneously.
For a successful multi-word attack, \emph{all} single-word attacks need to be successful.
For this experiment, we evaluate the attack for sentences with two, three, and four digits. 
For each set, we select 20 random audio files and aim to replace all the words with randomly chosen adversarial words.
As an example, the adversary might try to fool the \ac{ASR} system to recognize an utterance of \texttt{O89} as \texttt{762}.
We continue to use a limited-knowledge attacker that does not have access to the victim's training parameters and network architecture. 
We use the same setup as before and \netThreeLayer{} as the victim's network architecture.

Table~\ref{table:sentences} shows the attack statistics for sentences with different numbers of words. For reference, we repeat the results for the single-word attack in Table~\ref{table:sentences}.
The attack remains effective for longer sequences of words albeit with a decreased success rate.
Also, the attack uses more poisoned data to perform a multiple-digit replacement compared to a single-word replacement attack.


\begin{table}[t]
\addtolength{\tabcolsep}{-3.5pt}
\vspace{-0.5em}
\caption{Results for target sentences with different numbers of words. Note that the performance of the single-word attack is also presented as a reference.}
\vspace{-0.8em}

\centering
\smallskip
\small
\label{table:sentences}
\begin{tabular}{l|r r r r} 
\toprule
 & \multicolumn{4}{c}{\textbf{Number of Words}} \\ \cmidrule[0.6pt]{2-5}
  & \multicolumn{1}{c}{1} & \multicolumn{1}{c}{2} & \multicolumn{1}{c}{3} & \multicolumn{1}{c}{4} \\
\midrule
\textbf{Poison data length (seconds)}      & 25.44                & 46.17               & 63.85               & 89.68                       \\
\textbf{\# Poison data samples}               & 387.83               & 630.39               & 841.16             & 1,289.85                    \\
\midrule
\textbf{Clean test accuracy (\%)} & 98.04 & 97.84 & 97.67 & 97.75         \\
\midrule
\textbf{Attack success rate (\%)} & 86.7 & 75.0 & 60.0 & 60.0                     \\

\bottomrule
\end{tabular}
\vspace{-1.3em}
\end{table}

\subsection{Speech Commands Dataset}
\label{exp:speechcommands}
To further examine the practical feasibility of our attack, we evaluate \sys{} on a larger \ac{ASR} system.
To this end, we use the \emph{Speech Commands} corpus~\cite{warden2018speech} used for keyword spotting.
This dataset consists of 105,829 one-word utterances and contains 35 different words:
\begin{itemize}
    \item \emph{Digits} \texttt{ZERO}, ..., \texttt{NINE}
    \item \emph{Common words for IoT or robotics applications.} \texttt{YES}, \texttt{NO}, \texttt{UP}, \texttt{DOWN}, \texttt{LEFT}, \texttt{RIGHT}, \texttt{ON}, \texttt{OFF}, \texttt{STOP}, and \texttt{GO}
    \item \emph{Command words.} \texttt{FORWARD}, \texttt{FOLLOW}, \texttt{BACKWARD}, and \texttt{LEARN}.
    \item \emph{Auxiliary words.} \texttt{BED}, \texttt{BIRD}, \texttt{CAT}, \texttt{DOG}, \texttt{HAPPY}, \texttt{HOUSE}, \texttt{MARVIN}, \texttt{SHEILA}, \texttt{TREE}, \texttt{VISUAL}, and \texttt{WOW}.
\end{itemize}
%
For our poisoning attack trials, we randomly select 15 audio files and for each sample, we pick a random adversarial target. 

To fit this dataset, we use a larger neural network as well as a larger language model with 350 states. 
We use the \netThreeLayerPlus{} architecture for our surrogate networks, but with a larger output layer of size $350$ to contain all required phones of the extended language model.
As before, we use a fixed \ac{HMM} during the attack, which is trained in advance by training an \ac{ASR} system for $16$ epochs on the clean training set, of which the last epoch includes Viterbi training.
We use this surrogate \ac{HMM} at the beginning of each step of the attack to train four surrogate networks on the latest version of the poisoned dataset for 20 epochs with a batch size of 32. We verify that the training converges at 20 epochs.
We use the Adam~\cite{kingma2014adam} optimizer with a learning rate of $1e^{-4}$ for poison~crafting.

For the victim, we use a network architecture consisting of four hidden layers with 300, 200, 200, and 200 neurons, respectively.
The victim trains the \ac{ASR} system from scratch for 31 epochs, of which the eleventh epoch enables Viterbi training.
For the victim's training, a learning rate of $4e^{-4}$ and a batch size of 64 is used.

With a poison budget of \(\poisonbudget=0.02\), \sys{} achieves a success rate of 73.3\% while poisoning only 0.14\,\% of the training set (116.73 seconds of audio). 
Table~\ref{table:speechcommands} shows the attack performance for each example. We successfully poisoned 11 of the 15 trials. 
In general, we need to poison more and longer audio sequences with this extended dataset but the attack remains successful in most of the cases. 

\begin{table*}[t]
\centering
\small
\vspace{-0.7em}
\caption{Evaluation of \sys{} on the \emph{Speech Commands} dataset using 15 different random attack examples. The poison budget \(\poisonbudget\) is 0.02, and the attacker uses four surrogate networks to craft the poisoned data. On average, \sys{} uses 116.73 seconds of poisoned data (0.14\,\% of the training set). The total length of the training data is 84,054 seconds. The average SNRseg for poison frames is 4.14.}
\label{table:speechcommands}
\begin{tabular}{ll|cc|c|cc}
\toprule
\multicolumn{1}{c}{\textbf{Original}} & \multicolumn{1}{c|}{\textbf{Adversarial}} & \multicolumn{2}{c|}{\textbf{Poisoned data}} & \multicolumn{1}{c|}{\textbf{Poisoned frames}} &  \multicolumn{1}{c}{\textbf{Attack}} & \multicolumn{1}{c}{\textbf{Clean test}}  \\
\multicolumn{1}{c}{\textbf{word}} & \multicolumn{1}{c|}{\textbf{word}} & \multicolumn{1}{c}{\textbf{length (seconds)}} & \multicolumn{1}{c|}{\textbf{\# samples}} & \multicolumn{1}{c|}{\textbf{SNRseg}} & \multicolumn{1}{c}{\textbf{successful?}} & \multicolumn{1}{c}{\textbf{accuracy (\%)}}  \\
\midrule
\multicolumn{1}{l}{learn} & on & \multicolumn{1}{c}{~31.59} & \multicolumn{1}{c|}{~396} & \multicolumn{1}{c|}{~7.99} &  \cmark & 86.83  \\
\multicolumn{1}{l}{nine} & four & \multicolumn{1}{c}{156.71} & \multicolumn{1}{c|}{1,887}  & \multicolumn{1}{c|}{~7.49} &  \cmark & 87.07  \\
\multicolumn{1}{l}{three} & six & \multicolumn{1}{c}{124.71} & \multicolumn{1}{c|}{1,654} & \multicolumn{1}{c|}{-1.74} & \xmark & 87.16  \\
\multicolumn{1}{l}{six} & off & \multicolumn{1}{c}{~91.55} & \multicolumn{1}{c|}{1,057} & \multicolumn{1}{c|}{-0.63} &  \cmark & 86.98  \\
\multicolumn{1}{l}{yes} & go & \multicolumn{1}{c}{140.74} & \multicolumn{1}{c|}{1,493} & \multicolumn{1}{c|}{~7.75} &  \cmark & 86.90 \\
\multicolumn{1}{l}{six} & five & \multicolumn{1}{c}{128.36} & \multicolumn{1}{c|}{1,584} & \multicolumn{1}{c|}{~7.39} &  \cmark & 87.72  \\
\multicolumn{1}{l}{follow} & three & \multicolumn{1}{c}{~51.06} & \multicolumn{1}{c|}{~865} & \multicolumn{1}{c|}{~1.72} &  \xmark  & 87.39 \\
\multicolumn{1}{l}{four} & zero & \multicolumn{1}{c}{164.14} & \multicolumn{1}{c|}{2,012} & \multicolumn{1}{c|}{~8.37} &  \cmark & 86.99  \\
\multicolumn{1}{l}{follow} & two & \multicolumn{1}{c}{~45.35} & \multicolumn{1}{c|}{~549} & \multicolumn{1}{c|}{~3.74} &  \cmark & 86.79  \\
\multicolumn{1}{l}{four} & yes & \multicolumn{1}{c}{184.95} & \multicolumn{1}{c|}{2,153} & \multicolumn{1}{c|}{~4.06} &  \cmark & 87.35    \\
\multicolumn{1}{l}{six} & seven & \multicolumn{1}{c}{217.60} & \multicolumn{1}{c|}{2,412} & \multicolumn{1}{c|}{~4.07} &  \xmark & 87.35  \\
\multicolumn{1}{l}{one} & forward & \multicolumn{1}{c}{~80.66} & \multicolumn{1}{c|}{1,064} & \multicolumn{1}{c|}{~5.09} & \cmark & 85.86 \\
\multicolumn{1}{l}{four} & up & \multicolumn{1}{c}{150.78} & \multicolumn{1}{c|}{1,659} & \multicolumn{1}{c|}{-1.67} &  \cmark & 86.77  \\
\multicolumn{1}{l}{up} & off & \multicolumn{1}{c}{~79.65} & \multicolumn{1}{c|}{1,025} & \multicolumn{1}{c|}{~3.07} & \xmark & 86.67  \\
\multicolumn{1}{l}{one} & down & \multicolumn{1}{c}{~94.10} & \multicolumn{1}{c|}{1,256} & \multicolumn{1}{c|}{~5.33} &  \cmark & 87.12  \\
\bottomrule
\end{tabular}
\end{table*}

\subsection{Over-The-Air Attack}
\label{sec:results:overview}

\begin{table*}[t]
\addtolength{\tabcolsep}{-2.8pt}
\vspace{-0.5em}
\caption{\sys{}'s evaluation after the transmission in three simulated rooms, selected from related work~\cite{szoke2019building}, and one real physical room.
For the TIDIGITS dataset, the numbers are for the poison samples that are generated in Section~\ref{exp:limitedKnowledgeOfTrainingSet} for the 0\,\% overlap setting.
For the Speech Commands dataset, we use the poisoned data that \sys{} crafted in Section~\ref{exp:speechcommands}.
}
\vspace{-.5em}
\centering
\small
\begin{tabular}{cc|ll|llll|llll}
\toprule

\multicolumn{2}{c|}{} & \multicolumn{2}{c}{} & \multicolumn{4}{c}{\textbf{TIDIGITS}} & \multicolumn{4}{c}{\textbf{Speech Commands}} \\

\multicolumn{2}{c|}{\textbf{Room}} & \multicolumn{1}{c }{\textbf{Mic.}} & \multicolumn{1}{c|}{\textbf{Speaker}} & \multicolumn{4}{c|}{\textbf{Attack succ. rate (\%)}} & \multicolumn{4}{c}{\textbf{Attack succ. rate (\%)}} \\

\multicolumn{1}{c}{\textbf{Type}} & \multicolumn{1}{c|}{\textbf{Dim. (\(m^3\))}} & \multicolumn{1}{c}{\textbf{Position}} & \multicolumn{1}{c|}{\textbf{Position}} & \multicolumn{1}{c}{\textbf{RT=0.4}} & \multicolumn{1}{c}{\textbf{RT=0.6}} & \multicolumn{1}{c}{\textbf{RT=0.8}} & \multicolumn{1}{c|}{\textbf{RT=1}} &
\multicolumn{1}{c}{\textbf{RT=0.4}} & \multicolumn{1}{c}{\textbf{RT=0.6}} & \multicolumn{1}{c}{\textbf{RT=0.8}} & \multicolumn{1}{c}{\textbf{RT=1}} \\ \midrule
\multicolumn{1}{c}{Simulated} & \multicolumn{1}{c|}{\(10.7 \times 6.9 \times 2.6\)} & \(1.0 \times 4.5 \times 1.3\) & \(8.1 \times 3.3 \times 1.4\) & 53.33 & 46.67 & 36.67 & 33.33 & 20.00 & 20.00 & 26.67 & 20.00 \\
\multicolumn{1}{c}{Simulated} & \multicolumn{1}{c|}{\(~4.6 \times 6.9 \times 3.1\)} & \(3.8 \times 3.2 \times 1.2\) & \(3.8 \times 5.3 \times 1.0\) & 63.33 & 60.00 & 50.00 & 46.67 & 60.00 & 53.33 & 40.00 & 33.33 \\
\multicolumn{1}{c}{Simulated} & \multicolumn{1}{c|}{\(~7.5 \times 4.6 \times 3.1\)} & \(0.4 \times 0.9 \times 1.1\) & \(6.9 \times 1.9 \times 2.6\) & 73.33 & 60.00 & 56.67 & 56.67 & 46.67 & 46.67 & 40.00 & 40.00 \\
\midrule
\multicolumn{1}{c}{Physical} & \multicolumn{1}{c|}{\(~3.7 \times 3.4 \times 2.4\)} & \(1.7 \times 2.7 \times 1.2\) & \(2.1 \times 0.5 \times 0.8\) & \multicolumn{4}{c|}{73.33} & \multicolumn{4}{c}{33.33}\\


\bottomrule
\end{tabular}
\vspace{-1em}
\label{table:exp:overtheair}
\end{table*}

Prior work on audio adversarial examples~\cite{yakura2018robust,schonherr2020imperio} has often struggled in an over-the-air setting: During the transmission over the air, the audio signal is altered, which may affect the poisoning success. 
In this following, we study the effects of transmission over the air on our poisoning attack. 

%
First, we consider a simulated setting. To this end, we use the \emph{Python RIR Simulator} implementation~\cite{campbell-21-rirsimulator} and simulate the transmission in a room via a convolution with a \ac{RIR}~\cite{allen-1979-image}.
We evaluate the attack in three simulated rooms with the microphone and the speaker being positioned randomly. For each setting, we use four different reverberation times between 0.4--1.0 seconds.
Second, we evaluate the attack in a real physical room with an \emph{iPhone 13 Pro} microphone and a \emph{JBL GO} speaker.

We consider both datasets. For the TIDIGITS dataset, we use the poison samples that are generated in Section~\ref{exp:limitedKnowledgeOfTrainingSet} for the 0\,\% overlap setting. Consequently, the adversary does not know the victim's \ac{DNN} architecture and training parameters as well as the training set (except for the poisoned data).
Note that the victim uses \netThreeLayer{} in this evaluation. 
For the Speech Commands dataset, we use the same poisoned data as in Section~\ref{exp:speechcommands}.

Table~\ref{table:exp:overtheair} shows the results for different reverberation times (RT) in seconds, room dimensions, speaker and microphone positions. In addition, we also report the results for the physical room. For the TIDIGITS dataset, \sys{} maintains a success rate of 33.3-73.3\% across different room settings as opposed to the success rate of 86.7\% when feeding the input directly to the recognizer.
For the \emph{Speech Commands} dataset, \sys{} maintains an attack success rate of 20-60\% across different room settings as opposed to the success rate of 73.3\% when feeding the input directly to the recognizer.

\subsection{Transferability}
\label{exp:end-to-end}


In the previous sections, we focused on hybrid \ac{ASR} systems, and our results demonstrated that these are vulnerable to dataset poisoning attacks.
In this experiment, we consider the effect of the poisons for other \ac{ASR} architectures. In particular, a victim that uses an end-to-end \ac{ASR} system.

For this, we use an end-to-end system designed for the task of \emph{keyword spotting}~\cite{shan-18-attention, myer-18-efficient, arik-17-convolutional} on the \emph{Speech Commands} dataset based on SpeechBrain~\cite{speechbrain}.\footnote{Recipe: \url{https://github.com/speechbrain/speechbrain/tree/develop/recipes/Google-speech-commands}} 
This \ac{ASR} system has a total of 4,494,777 trainable parameters.
For reference, the hybrid system that we evaluated in Section~\ref{exp:speechcommands} has a total of 265,295 trainable parameters, which is 0.06 times less than the end-to-end~system.

We use the same poison samples generated in Section~\ref{exp:speechcommands} to attack hybrid \ac{ASR} systems. For each of the 11 successful attack examples, we evaluate the victim's end-to-end system by training it on the poisoned datasets.
We observe that the attack fools the victim's end-to-end system for four examples, showing a transferability rate of 36.4\%.
The test accuracy for the poisoned models is on average at 95.06\%.

\begin{table}[t]
\centering
\smallskip
\small
\vspace{-.5em}
\caption{Results for different levels of \psycho{} filtering~\(\hearingOffset\) (poison budget~\(\poisonbudget\) is set to 0.005). 
}
\vspace{-0.5em}

\begin{tabular}{c|c|cc}
\toprule
 & \multicolumn{1}{c|}{\textbf{Poisoned frames}} & \multicolumn{1}{c|}{\textbf{Attack succ.}} & \multicolumn{1}{c}{\textbf{Clean test}} \\
\textbf{\(\bm{\hearingOffset}\) (dB)} & \multicolumn{1}{c|}{\textbf{SNRseg}} & \multicolumn{1}{c|}{\textbf{rate (\%)}} & \multicolumn{1}{c}{\textbf{acc. (\%)}} \\
 \midrule
 20 & 4.61 & ~0.0 & 97.80 \\
 30 & 4.25 & 43.3 & 97.80 \\
 40 & 3.54 & 66.7 & 97.81 \\
 50 & 4.13 & 80.0 & 97.80 \\
 \texttt{NONE} & 2.17 & 86.7 & 97.84 \\
 \bottomrule
\end{tabular}
\vspace{-1.5em}
\label{table:exp:hearingOffsets}
\end{table}
\subsection{User Study}
\label{sec:userStudy}

To evaluate the human perception of our poison samples, we conduct a listening test, where we ask participants to transcribe utterances of the poisoned data.
Furthermore, in this section, we additionally consider \psycho{} modeling~\cite{zwicker2013psychoacoustics, schonherr2018adversarial} as a mechanism to limit the perceptible perturbations introduced by the attack.



\subsubsection{\Psycho{} Modeling}
To make poisons less conspicuous, we can utilize \psycho{} modeling to limit audible distortions.
Recent attacks against \ac{ASR}~\cite{schonherr2018adversarial, qin2019imperceptible} proposed \psycho{} hiding as a method to create less perceptible adversarial noise.
To identify inaudible ranges, these attacks use dynamic hearing thresholds, which describe the masking effects in human perception that arise as a function of the interactions between different co-occurring acoustic frequencies.
We implement \psycho{} hiding similar to what is described by Sch\"onherr~\etal~\cite{schonherr2018adversarial}. Appendix~\ref{app:psycho} elaborates in detail how we employ \psycho{}~filtering.

We evaluate \sys{} for varying degrees of \psycho{} filtering, controlled through margin~\(\hearingOffset\) (in dB) that allows the attack to surpass the hearing thresholds.
The higher \(\hearingOffset{}\), the more audible noise is allowed.
As shown by Table~\ref{table:exp:hearingOffsets}, enabling the \psycho{} hiding decreases the attack success rate, while the SNRseg of poisoned frames improves.
The case without enforcing hearing thresholds is denoted as \texttt{NONE}.
Note that the choice of poison samples and frames does not depend on the margin~\(\hearingOffset\);
that is, the average length of the poisoned data is always 25.44s in Table~\ref{table:exp:hearingOffsets}.

\subsubsection{Transcription Test}
For the study, we randomly selected 20 poison samples from 12 successful attack examples, both when the \psycho{} hiding was disabled and for \(\hearingOffset\!=\!30\)\,dB, which resulted in a pool of 480 poison samples.
For verification, participants also transcribed five hidden clean samples.

We asked 23 English speakers to transcribe a random subset of utterances. The participants were not informed if a sample has been modified or if it represents a clean sample. 
On average, each user transcribed 40 poison samples. 
For each attack example, we report the ratio of the poison samples that are transcribed into their original label.

When the \psycho{} hiding is disabled, 87.1\,\% of the poison samples were transcribed into their original labels. 
On the other hand, for \(\hearingOffset\!=\!30\)\,dB, 85.0\,\% of the poison samples were transcribed into their original labels. 
These results show that even though enforcing hearing thresholds of \(\hearingOffset=30\)\,dB improves the SNRseg values of the poisoned frames (from 2.17 to 4.25, see Table~\ref{table:exp:hearingOffsets}), the performance of the transcription test is not improved. 

The results of this feasibility study also indicate that the poisoned data generated by \sys{} contain samples that can be considered as clean-label samples. 
Such a study has often been missing in prior works, and as noted by Schwarzschild et al.~\cite{schwarzschild2021just}, most current attacks in the visual domain produce easily visible artifacts and distortions.  

\section{Discussion} 
\label{sec:desc}
Next, we expand our analysis of \sys{} by providing insights into our results. We will also summarize the results and discuss major findings and limitations.

%

\subsection{Attack Parameters}
Here, we discuss the impact of \sys{}'s parameters on the attack success rate.

\subsubsection{Poison Budget \& Surrogate Models}
Using a larger poison budget \(\poisonbudget\) increases the number of poisoned files (and frames). However, we show that beyond a poison budget of 0.005, the attack success does not further improve (see Table~\ref{table:exp:diffpoisonbudget}), and, therefore, more poison samples are not necessarily required for the attack. 
The same can be observed for the number of surrogate models; using more surrogate models does not necessarily increase the attack's success (see Table~\ref{table:exp:diffnumsurroagtenetworks}).

\subsubsection{Target Selection}
In Section~\ref{exp:multiword}, we show that \sys{} is not limited to the replacement of single words; it can successfully replace all the words with the intended adversarial words. Consequently, an attacker has full control of the output of the target, and arbitrary transcriptions can be chosen.
%
This is further supported in our experiments with the Speech Commands dataset, where we show that \sys{} scales to \ac{ASR} systems with a larger vocabulary. 

To further understand how the number of \ac{HMM} states of the target word affects the success rate of \sys{}, we consider our single-word replacement attack in Section~\ref{exp:limitedKnowledgeAdversary} on the TIDIGITS dataset.
We conducted this experiment over 30 trials, which we divide here into three different categories: (1) In 11 trials, the target word has more HMM states than the original word, (2) in 7 trials, the target word and the original word have the same number of HMM states, and (3) in 12 trials the target word has less HMM states than the original word. For the results presented in Table~\ref{table:diffVictimSettingsSmallTable} (last column), the attack fails on two, one, and two trials, respectively, in these three types of trials, showing that the difference between the number of \ac{HMM} states of the target and original word does not affect the success rate of the attack.

\subsubsection{Sequence Selection}
To quantify the effect of the sequence selection on the attack success rate, we repeat the experiment from Section~\ref{exp:limitedKnowledgeAdversary} (Table~\ref{table:diffVictimSettingsSmallTable}). Instead of choosing the target sequences based on the frequency analysis (explained in Section~\ref{sec:method:overview}), we now \emph{randomly} select the target sequence. We require that the sequence has to be in ascending order (e.g., for a target sequence like [92, 92, 91, 91, 93, 93] the language model can otherwise not return a valid word). In this experiment, we observe a drop in the attack success rate by 23.33 percentage points (from 83.33\% to 60.0\%).

\subsection{Clean Test Accuracy}
In our evaluation, we always use the entire test dataset to calculate the clean accuracy using the \emph{edit distance} between the ground-truth label and the predicted transcription.
Here, we aim to understand how the attack affects the recognition of the target word in isolation. We use the results presented in Section~\ref{exp:limitedKnowledgeAdversary} for the following measurements:
\begin{itemize}
    \item For each digit, we only consider the test audio files that contain the digit to calculate the number of errors (I + S + D, Section~\ref{sec:exp}). On average over 30 trials, the total number of errors for the target and original digits are 93 and 95 words, respectively, while the number of errors for the other digits is 111 words.
    \item For each digit, we consider the test audio files that do not contain the digit. For these files, we count how often the model’s transcription (mistakenly) contains the digit. On average over 30 trials, for 9.97 utterances, the model mistakenly recognizes the target digit. For the original digit, this value is 8.97, while for the other digits, this value is 10.26 on average.
\end{itemize}

\subsection{Practical Considerations}

In the following, we elaborate on the practical aspects of our attack and reflect on its implications and limitations.

\subsubsection{Clean-Label Poison Utterances}
In the listening test, we verify that \sys{} is able to generate clean-label poison samples. We ask participants to transcribe poisoned audio samples and on average, more than 85\% of the poison samples were transcribed into their original labels, showing that even manual verification of training data would not be effective to prevent audio poisoning attacks. 

Furthermore, in privacy-preserving \emph{federated learning} scenarios, where the training data and the training is decentralized, a party can easily compromise the training data~\cite{tolpegin-20-data}. 
Here, the poison samples are not constrained to clean-label data points, as the victim has no access to the training data, while the attacker has full control of their data.
Additionally, our limited-knowledge experiments have shown that controlling only parts of the training process and training data---as would be the case in a federated learning scenario---is very effective.

\subsubsection{Limited Vocabulary} 
We showed our attack is successful on two datasets, TIDIGITS and Speech Commands, of which the latter is ten times bigger than the former. We argue that our results show that data poisoning attacks against \ac{ASR} systems are a viable threat that needs to be considered by researchers working on \ac{ASR} systems. Based on our foundations, we hope that future work will improve the scalability of our attack and include larger datasets in their evaluation and develop more robust \ac{ASR} systems that are resistant to data poisoning attacks.

\subsubsection{Fine-Tuning}
Although hybrid \ac{ASR} systems are typically trained from scratch, we now want to expand our evaluation and also consider a fine-tuning scenario.
For this, we use the poisoned data generated for the most restricted adversary (Table~\ref{table:exp:diffVictimTrainingSets}). 
That is, the adversary’s training set is the “Split 1” subset. For the victim’s model, we divide the “Split 2” subset into two parts of equal size (each with 28 speakers). 
The first part is the training set and contains only clean data. 
The second part, which is the fine-tuning set, is poisoned. 
On average, over the same 30 trials, we observe an attack success rate of 63.33\% (83.33\% for the from-scratch training scenario).
For training and fine-tuning, we used a learning rate of 1e-4 and 5e-5, respectively.

\subsubsection{Over-the-Air} In Section~\ref{sec:results:overview}, we demonstrate that \sys{} is also successful if the targeted audio signal is played over the air in simulated and physical rooms of different sizes. This shows the general robustness of our attack and that the poison samples also remain effective after a transmission's alterations. Notably, the attack is generic in the sense that the properties of the room need not be known beforehand. 

\subsubsection{Transferability To End-To-End Keyword Spotting}
To verify the practicality of \sys{} in the real world, we evaluate the poisoned data generated by the attack against an end-to-end \ac{ASR} system, designed specifically for the task of keyword spotting on the Speech Commands dataset.
Our results in Section~\ref{exp:end-to-end} show that although the poison samples of \sys{} are not crafted for end-to-end systems, they remain viable and can be a potential threat to such systems.

\subsubsection{Hearing Thresholds}
Hearing thresholds have shown to be effective for adversarial examples, however, in the case of poisoning, we observe that their effect is less distinct. One main reason may be that in contrast to adversarial examples, where the complete file is modified, our modifications for the poison utterances are limited to short~sequences.

\section{Related Work}
\label{sec:relwork}
In the following, we discuss related work on attacks against machine learning and \ac{ASR} systems.

\subsubsection{Adversarial Examples} Adversarial examples are carefully crafted inputs that are perturbed by adding imperceptible noise to fool a machine learning model~\cite{szegedy2013intriguing, biggio2013evasion}.
Such perturbations are calculated using the gradients of an optimization problem that is defined on the victim network, or surrogate networks, if the victim network is unknown.
Initial work on adversarial attacks focused on the space of images~\cite{goodfellow2014explaining, biggio2013evasion}.
Later, similar evasion attacks were shown to exist in the audio domain, where generating adversarial examples is more challenging due to time dependencies that exist in the \ac{ASR} systems
~\cite{vaidya2015cocaine, carlini2018audio, schonherr2018adversarial, schonherr2020imperio}.


\subsubsection{Backdoor Attacks}
For a backdoor attack, an adversary manipulates the victim model by imprinting training samples with a specific pattern (\textit{trigger}) and the target label to train the model to become sensitive to this pattern~\cite{gu2017badnets}.
During inference, the attacker can then cause a misclassification by injecting the trigger into \emph{any} input example. By using ultrasonic triggers, the feasibility of such an attack against \ac{ASR} was recently demonstrated in a technical report by Koffas \etal \cite{koffas2021can}. In contrast to our work and similar to evasion attacks, however, backdoor attacks require the modification of test samples during inference, which is not always applicable in real-world scenarios.

\subsubsection{Training-Time Poisoning Attacks}
Closest to our work are \emph{training-time poisoning attacks}~\cite{shafahi2018poison, zhu2019transferable, aghakhani2020bullseye, huang2020metapoison, geiping2020witches} against image classification, wherein the adversary crafts poison images---with \emph{no} control over the labeling process---to achieve the system's misbehavior for specific target inputs.
There exist major limitations with these attacks, which hinder their application to \ac{ASR} systems. 
First, these attacks focus on transfer learning, which is not a common training practice for speech recognition; \ac{ASR} systems are typically trained from scratch.
Second, they assume that the victim does not use dropout during the fine-tuning process, while dropout is often enabled in training neural networks from scratch.
Furthermore, unlike image classification, the recognition process of \ac{ASR} is based on time series signals (\ie the waveform audio signal).
Consequently, these attacks cannot directly be applied to speech-based systems. 

\subsubsection{Countermeasures}
Although several automated defenses have been proposed~\cite{peri2020deep, liu2018fine, chacon2019deep}, they can typically be evaded by an adaptive attacker~\cite{koh2018stronger, schwarzschild2021just}. 
One line of possible defenses focus on poison detection and removing them from the train set.
This usually happens by employing some neighborhood conformity tests or outlier detection, either on the data itself or in the latent space~\cite{peri2020deep}. This type of detection, however, requires access to the training data, which is not always given (\eg in a federated learning setting). Most recent defenses also consider retrospective countermeasures like forensic-inspired approaches~\cite{shan-22-forensics}. 
Their strategy is to detect the origin of the poisoned data \emph{after} a successful attack, and, therefore, cannot prevent harm beforehand.
%

Other defenses try to detect poisoned models~\cite{peri2020deep, liu2018fine, chacon2019deep}. However, these sanitization-based defenses may be easily leveraged by an attacker who is aware of the specific defense mechanism, as they are attack-specific~\cite{koh2018stronger, schwarzschild2021just}. More importantly, most defenses require clean reference data to sanitize the training data. The distribution of such clean data needs to be close to the distribution of the training data, which is often not realistic.

\section{Conclusions}
\label{sec:conc}

In this paper, we present \sys{}, the first training-time poisoning attack against speech recognition. 
In a series of experiments, we demonstrate \sys{}'s efficacy and evaluate the attack under different attack settings and for various attack parameters. We test single and multi-word replacement attacks and investigate the effect of an enlarged language model.
The attack remains viable in an over-the-air scenario, with limited knowledge about the victim model, and transfers between different speech recognition architectures.
Finally, we verify with a user study that the majority of poison samples are clean-label, which renders a manual verification of the training data ineffective.
In summary, we show with \sys{} that data poisoning of \ac{ASR} systems poses a real threat that needs to be considered.

\section*{Acknowledgments}
We would like to thank our reviewers for their valuable comments and input to improve our paper.
This material is based upon work partially supported by NSF under Award \#CNS-2107101 and by a gift from Intel, Corp. Any opinions, findings, and conclusions or recommendations expressed in this publication are those of the author(s) and do not necessarily reflect the views of NSF or Intel.
Moreover, this work was funded by the Deutsche Forschungsgemeinschaft (DFG, German Research Foundation) under Germany's Excellence Strategy -- EXC 2092 CASA -- 390781972.



\bibliographystyle{plain}
\bibliography{strings, main}


\section{Appendix}


\subsection{\Psycho{} Modeling}
\label{app:psycho}

Recent adversarial attacks against \ac{ASR} systems~\cite{schonherr2018adversarial, qin2019imperceptible} use \psycho{} hearing thresholds to hide modifications of the input audio signal within inaudible ranges.
By using hearing thresholds, we can limit audible distortions.
These thresholds define how dependencies between certain frequencies can mask, i.e., make inaudible, parts of an audio signal.
In essence, we guide \sys{} to hide malicious noise in these inaudible parts.
At each step of the poison crafting, we scale the gradients of the poison audio signal (calculated via minimizing Equation~\ref{eq:BPLoss}) with scaling factors that limit audible distortions.
Since human thresholds alone are tight, the scaling factors are allowed for differing from the thresholds by a margin of \(\bm{\hearingOffset{}}\) (in dB).
The higher \(\bm{\hearingOffset{}}\), the more audible noise is allowed to be added by the attack.

In the following, we discuss how we compute the scaling factors.
First, we compute the power spectrum of the difference \textbf{\diffspect{}} between the poison signal spectrum \(\bm{\poisonspect{}}\) and the original signal spectrum \textbf{\origspect{}} for all times \(t\) and frequencies \(q\) as the following:
\begin{align}
\diffspect{}(t,q) = 20 \times \textrm{log}_{10} \frac{|\poisonspect{}(t,q)-\origspect{}(t,q)|}{\textrm{max}_{t,q}(|\origspect{}|)}, \forall t,q.\notag
\end{align}

Then we compute the audible difference (in dB) for all times \(t\) and frequencies \(q\) via
\begin{align}
\audibleDiff{}(t,q)=
\textrm{\diffspect{}} - \textrm{\hearingThresholds{}},\notag
\end{align}
where \textbf{\hearingThresholds{}} is the computed human hearing thresholds based on the \psycho{} model of MPEG-1~\cite{iso-93-part3}.
Since the thresholds~\hearingThresholds{} are tight, we allow \sys{} to differ from the hearing thresholds by a margin of~\(\bm{\hearingOffset{}}\) (in dB).
In particular, we calculate the matrix~\(\audibleDiffMinZero\) for all times \(t\) and frequencies \(q\) as

\vspace{-2em}
\begin{align}
\audibleDiffMinZero{}(t,q)=
    \begin{cases}
        \hearingThresholds{}(t,q) + \hearingOffset{} - \diffspect{}(t,q) & \textrm{if} \; \; \hearingThresholds{}(t,q) + \hearingOffset{} \geq \diffspect{}(t,q)\\
        0 & \textrm{else}\notag
    \end{cases}
\end{align}
where we clip the negative values to zero for the time-frequency bins that cross the thresholds \(\textrm{\hearingThresholds{}}\,+\,\hearingOffset{}\).
We then normalize the matrix \(\audibleDiffMinZero{}\) to values between zero and one via
\begin{align}
    \audibleDiffNormalized{}(t,q)=\frac{\audibleDiffMinZero{}(t,q)-\textrm{min}_{t,q}(\audibleDiffMinZero{})}{\textrm{max}_{t,q}(\audibleDiffMinZero{})-\textrm{min}_{t,q}(\audibleDiffMinZero{})}, \forall t, q.\notag
\end{align}
We also compute a fixed scaling factor by normalizing the hearing thresholds \hearingThresholds{} to values between zero and one via
\begin{align}
    \hearingThresholdsNormalized{}(t,q)=\frac{\hearingThresholds{}(t,q)-\textrm{min}_{t,q}(\hearingThresholds{})}{\textrm{max}_{t,q}(\hearingThresholds{})-\textrm{min}_{t,q}(\hearingThresholds{})}, \forall t, q.\notag
\end{align}
Putting the scaling factors \(\audibleDiffNormalized{}\) and \(\hat{\textrm{H}}\) together, the gradient of \(\nabla X\) computed via Equation~\ref{eq:BPLoss} will be scaled as the following
\begin{align}
    \nabla X_{(t,q)}:=\nabla X_{(t,q)} \cdot  \audibleDiffNormalized{}(t,q) \cdot  \hearingThresholdsNormalized{}(t,q), \forall t,q.\notag
\end{align}
This scaling happens between the \ac{DFT} and the magnitude step in the computational graph.

\end{document}